\documentclass[10pt]{iopart}

\usepackage{hyperref}

\usepackage{framed,multirow}
\usepackage{amsmath} %added for maths environments (equation and align)
\usepackage{amssymb}
\usepackage{bm} %bold symbols by using \bm
\usepackage{mathtools} %added for \xrightleftharpoons
\usepackage{upgreek} %upright bold greek letters
\usepackage[fleqn]{nccmath}
\usepackage[capitalise, nameinlink]{cleveref}
\crefname{appendix}{}{}

\usepackage{subcaption} %allowing for subcaptions and subfigures
\usepackage{algorithm} %added for algorithm box
\usepackage{algpseudocode}

\usepackage{placeins}

\usepackage[a4paper, margin=1in]{geometry}

\renewcommand{\thetable}{\Roman{table}}  %tables using roman numerals

\newcommand\citep[1]{\cite{#1}} %citation style and using superscript
\usepackage[superscript,biblabel]{cite}

\begin{document}

\title{Stabilising effects of lumped integration schemes for the simulation of metal-electrolyte reactions}

\author{Tim Hageman, Emilio Martínez-Pañeda}
\ead{e.martinez-paneda@imperial.ac.uk}

\address{Department of Civil and Environmental Engineering, Imperial College London, London SW7 2AZ, UK}

\begin{abstract}
Computational modelling of metal-electrolyte reactions is central to the understanding and prediction of a wide range of physical phenomena, yet this is often challenging owing to the presence of numerical oscillations that arise due to dissimilar reaction rates. The ingress of hydrogen into metals is a paradigmatic example of a technologically-relevant phenomenon whose simulation is compromised by the stiffness of the reaction terms, as reaction rates vary over orders of magnitude and this significantly limits the time increment size. In this work, we present a lumped integration scheme for electro-chemical interface reactions that does not suffer from numerical oscillations. The scheme integrates the reactions in a consistent manner, while it also decouples neighbouring nodes and allows for larger time increments to be used without oscillations or convergence issues. The stability and potential of our scheme is demonstrated by simulating hydrogen ingress over a wide range of reaction rate constants and environmental conditions. While previous hydrogen uptake predictions were limited to time scales of minutes, the present lumped integration scheme enables conducting simulations over tens of years, allowing us to reach steady state conditions and quantify hydrogen ingress for time scales relevant to practical applications.
% Highlights (max 85char per highlight, graphical abstract optional)
% A lumped numerical integration scheme is presented for (electro-) chemical reactions
% This scheme surpresses non-physical oscilations, greatly enhancing stability
% Compared to Gauss integration, larger ranges of reaction constants can be simulated
% Application to hydrogen embrittlement is detailed, highlighting its capabilities
% The described scheme makes simulation over all relevant time scales achievable
\end{abstract}

\vspace{2pc}
\noindent{\it Keywords}: electrochemistry, finite element method, oscillations, stabilisation, hydrogen embrittlement, lumped integration

\submitto{Journal of the Electrochemical Society}

%\linenumbers

\section*{Introduction}
\label{sec:intro}

An accurate estimation of metal-electrolyte reactions is key to a number of disciplines, from corrosion to catalysis. However, this is often hindered by the numerical instability problems that arise because of the differences in reaction rate magnitudes. One area where this issue is particularly pressing is the field of \textit{hydrogen embrittlement} - the fracture toughness and ductility of metallic materials is very sensitive to the dissolved hydrogen content and thus being able to quantify hydrogen uptake into a metal is paramount \citep{Gangloff2008, Martinez-Paneda2021}. A vast literature has been devoted to the development of models aimed at predicting hydrogen-assisted failures (see \cite{Yu2016, Nagao2018, Martinez-Paneda2018, Anand2019, Wu2020, Golahmar2021} and Refs. therein), but these take as input the hydrogen concentration associated with a given environment, which is generally an unknown quantity (outside of gaseous environments and steady state conditions). Very often, simulations assume a constant hydrogen concentration at the boundaries of the domain \citep{Moriconi2014, Duda2018, Colombo2020,Isfandbod2021}. However, this is unable to account for the role of mechanical stresses and surface absorption rates. It is slightly more accurate to prescribe a chemical potential-based boundary condition, allowing the interactions between mechanical strains in the metal and hydrogen ingress to be captured \cite{DiLeo2013, Martinez-Paneda2016, Diaz2016b, Elmukashfi2020, Fernandez-Sousa2020}. A more rigorous description is given by the generalised boundary conditions proposed by Turnbull and co-workers \cite{Turnbull1996,Martinez-Paneda2020}, whereby a hydrogen absorption flux is defined based on a fixed pH and overpotential. However, while these two approaches incorporate into the boundary conditions information about the environment, they carry the assumption that the environment remains unaltered. Very recently, an electro-chemo-mechanical framework has been developed to capture the metal-electrolyte interactions and ionic transport within the electrolyte \citep{Hageman2022}. It was shown that the hydrogen evolution and corrosion reactions have a significant impact on the pH of the electrolyte, which in turn strongly influences the hydrogen uptake within the metal. However, during these simulations stability issues were encountered originating from the high reaction rates at the metal surface and within the electrolyte. These issues manifested themselves as severe oscillations and imposed the need for small time increments to retain a stable simulation, precluding predictions over time scales relevant to hydrogen embrittlement.\\ 

While these stability issues are yet to be addressed in the area of electrochemistry, inspiration can be taken from other disciplines. For example, lumped integration schemes have been used in solid mechanics to prevent oscillations arising at interfaces when penalty approaches are used to prevent interpenetration \citep{Schellekens1993,Vignollet2015}. And in geomechanics, the use of lumped integration for pressure capacity terms has shown to reduce oscillations in fluid pressure and fluid flux when simulating hydraulic fractures \citep{Li2018,Hageman2019b,Hageman2020a}. Based on these promising results in other fields, we here propose to use a lumped integration scheme to address numerical instabilities in metal-electrolyte reactions. We choose to demonstrate this new computational paradigm in the area of electrochemistry by simulating the reactions involved in the uptake of hydrogen into metals. However, it should be noted that the lumped integration is applied to surface reaction and volume reaction terms, and these terms are not solely limited to hydrogen ingress. For instance, the removal of oscillations is shown for the water auto-ionisation reaction within the context of hydrogen absorption, but the same approach could be used for any water-based electrolyte. Similarly, while the electrolyte-metal interactions considered are particularly relevant for hydrogen uptake, the proposed method is applicable to any system where ``fast" surface reactions might occur, such as corrosion under anodic potentials and Li-Ion battery operation.\citep{Tsuyuki2018, Zhao2022} As such, the implications of the results are much wider and relevant to any simulation of fast reactions occurring within electrolytes or at surfaces.\\

The remainder of the manuscript is organised as follows. First, the specific choices of governing equations adopted are provided, and the lumped integration scheme is described. The ability of the lumped integration scheme in opening new modelling horizons for hydrogen ingress is addressed next, after which a representative case study is used to compare the results obtained with this new lumped integration scheme to those obtained with a conventional Gauss integration approach, emphasising the stabilising capabilities of the former. The stability of the integration scheme presented is further explored and demonstrated through simulations spanning a wide range of reaction constants and environmental conditions. These reveal that the scheme is stable for all realistic parameters and capable of pioneeringly providing hydrogen uptake predictions over technologically-relevant time scales. Finally, we present results for a more complex three-dimensional case, a tensile rod undergoing hydrogen uptake while subjected to various mechanical loads and electric potentials. This case study not only provides further evidence of the ability of the present scheme to simulate complex cases over large time scales, but it is also exploited to bring new insight - mapping the relationship between applied strain, imposed electric potential, and the resulting hydrogen uptake, as well as the time required for this uptake to occur. The paper concludes with a summary of the main findings. 

\FloatBarrier
\section*{Governing equations}
\label{sec:governing_eq}
\begin{figure}
    \centering
    \includegraphics{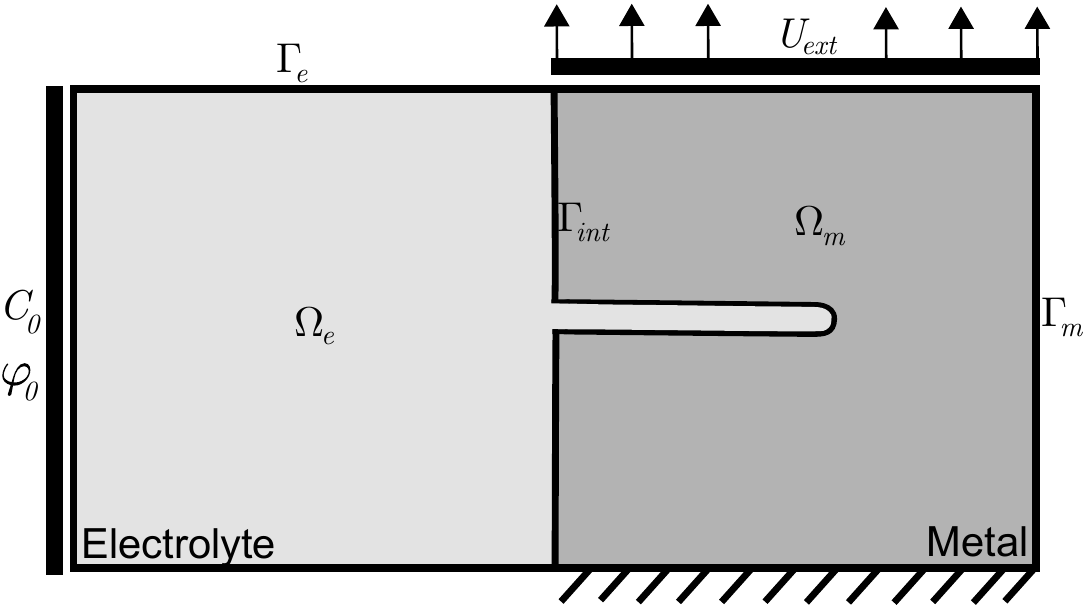}
    \caption{Overview of domain considered, composed of the electrolyte sub-domain $\Omega_e$ with external boundary $\Gamma_e$, metal sub-domain $\Omega_m$ with external boundary $\Gamma_m$, and the electrolyte-metal interface $\Gamma_{int}$.}
    \label{fig:domain_overview}
\end{figure}

Throughout this paper, we consider the domain $\Omega$ of the type shown in \cref{fig:domain_overview}. This domain consists of an electrolyte region $\Omega_e$, characterised by its primary fields the ionic species concentrations $C_\pi$ and the electrolyte potential $\varphi$, and a neighbouring metal region $\Omega_m$ described by the displacement field vector $\mathbf{u}$ and the lattice hydrogen concentration $C_L$. Separating these two sub-domains is the metal-electrolyte interface $\Gamma_{int}$ which is included through the hydrogen surface coverage $\theta_{ads}=C_{ads}/N_{ads}$, where $N_{ads}$ is the number of surface adsorption sites and $C_{ads}$ is the surface concentration. The metal domain contains a crack, as near crack tips is where the critical content of hydrogen triggering embrittlement is attained.\\

Regarding notation and units. For consistency between the metal and electrolyte domains, the ionic concentrations in the electrolyte and the hydrogen concentration within the metal are expressed in the SI units $\mathrm{mol}/\mathrm{m}^3$. Consequently, for all constants related to these quantities, such as the reaction constants, their units follow accordingly; e.g., the water auto-ionization constant is given by $10^{-8}\;(\mathrm{mol}/\mathrm{m^3})^2$ (as opposed to the more common terminology of $10^{-14}\;(\mathrm{mol}/\mathrm{L})^2$). As a reference point for the electrolyte potential, the standard hydrogen electrode is used, placing the equilibrium constant of the hydrogen evolution reaction at $0\;\mathrm{V}_{\mathrm{SHE}}$. Finally, we use lightface italic letters for scalars, e.g. $C_L$, upright bold letters for vectors, e.g. $\mathbf{u}$, and bold italic letters, such as $\bm{\sigma}$ or $\bm{B}$, for second and higher order tensors.

\subsection*{Electrolyte sub-domain}
For the electrolyte sub-domain, we consider the following ionic species: $\mathrm{H^+}$ and $\mathrm{OH}^-$ to represent a water-like electrolyte, $\mathrm{Na}^+$ and $\mathrm{Cl}^-$ to approximate seawater, and $\mathrm{Fe}^{2+}$ and $\mathrm{FeOH}^+$ to track the products produced by corrosion. We assume that the solubility of any gas phases produced by reactions is negligible, and that these gases disappear from the domain immediately.\\ 

The transport of ionic species inside the electrolyte is characterised by the Nernst-Planck mass balance, which describes changes in concentration based on advection, diffusion due to gradients in concentration, and electromigration due to gradients in electric potential. Thus, for a volume reaction rate $R_\pi$, charge of the ionic species $z_\pi$, diffusivity $D_\pi$, Faraday constant $F$, reference temperature $T$, and gas constant $R$, the Nernst-Planck equation reads
\begin{equation}
    \dot{C}_{\pi}+(\mathbf{v}\cdot\bm{\nabla})c_\pi+\bm{\nabla}\cdot\left(-D_\pi \bm{\nabla}C_\pi\right) + \frac{z_\pi F}{RT} \bm{\nabla} \cdot \left(-D_\pi C_\pi \bm{\nabla} \varphi\right) +R_\pi = 0 \label{eq:nernstplanck}
\end{equation}
where $\mathbf{v}$ represents the fluid velocity of the electrolyte, which we will assume equal to zero in the remainder to solely focus on reaction-induced convergence issues and oscillations, instead of including advection-based transport and the stability issues associated with it \citep{Hughes1995, Masud2004, Hughes2005, Hauke2007, Ropp2009, Nadukandi2010, Bauer2012}.\\ 

In addition to the mass balance, we require the local electric charge to be conserved. Several popular formulations for this charge-conservation include tracking the total current flow and requiring it to be divergence-free, $ \bm{\nabla}\cdot (\sum z_\pi \mathbf{j}_\pi)=0$ \citep{Pillay1993, Duddu2016}, or using Gauss law to describe the charge conservation within the electrolyte and the electron double-layer, $\bm{\nabla}^2\varphi=-F/\varepsilon\;\sum z_\pi C_\pi$ \citep{Sarkar2011, Wang2021}. However, since this double layer is extremely thin and infeasible to account for through the discretisation of the electrolyte domain \citep{Feldberg2000}, we have chosen to adopt the electroneutrality condition because of its simplicity; such that
\begin{equation}
    \sum_\pi z_\pi C_\pi = 0
    \label{eq:electroneutrality}
\end{equation}
with this equation inducing changes in electrolyte potential to enforce the redistribution of species through the electrolyte potential-based diffusion term from \cref{eq:nernstplanck}.\\ 

In addition to these conservation laws, we consider several reactions to occur within the electrolyte. Since a water-like electrolyte is considered, the water auto-ionisation reaction is relevant \citep{Wrubel2020}:
\begin{equation}
    \mathrm{H}_2\mathrm{O} \xrightleftharpoons[k_{w}']{k_{w}} \mathrm{H}^+ + \mathrm{OH}^- \label{react:water}
\end{equation}
and it is implemented through the reaction term $R_\pi$ from \cref{eq:nernstplanck} as:
\begin{equation}
    R_{\mathrm{H}^+,w}=R_{\mathrm{OH}^-} = k_{w}C_{\mathrm{H}_2\mathrm{O}} - k_{w}'C_{\mathrm{H}^+}C_{\mathrm{OH}^-}  = k_{eq} \left(K_w-C_{\mathrm{H}^+} C_{\mathrm{OH}^-} \right) \label{eq:water_react}
\end{equation}
Here, we assume equilibrium between the forward reaction rate, governed through $k_w$, and the backward reaction rate $k_w'$. This allows us to rewrite the reaction rate in terms of the equilibrium constant $K_w=10^{-8}\;\mathrm{mol}^2/\mathrm{m}^6$ and a penalty term-like constant $k_{eq}$, where this penalty constant is chosen to be high enough to enforce the equilibrium reaction to be fulfilled at all times. An alternative to this penalty approach is using the equilibrium reaction to eliminate $C_{\mathrm{OH}^-}$ from the governing equations. However, this removes the boundary influx terms and would therefore complicate applying (electro-) chemical reactions involving this species on the metal-electrolyte boundary. It is furthermore not an approaches that can be generalised for the case of multiple equilibrium reactions. Therefore, we have chosen to use the penalty approach to enforce the equilibrium and resolve the complications with regards to stability and oscillations, allowing for an easy and straightforward integration of bulk and surface reactions involving an arbitrary amount of reaction and involved reaction species.\\

For the corrosion products, we consider the following reaction:
\begin{equation}
    \mathrm{Fe}^{2+} + \mathrm{H}_2\mathrm{O} \xrightleftharpoons[k_{fe}']{k_{fe}} \mathrm{FeOH}^+ + \mathrm{H}^+ \label{react:fe_feoh}
\end{equation}
which in turn can react further through:
\begin{equation}
    \mathrm{FeOH}^{+} + \mathrm{H}_2\mathrm{O} \xrightharpoonup{k_{feoh}} \mathrm{Fe}(\mathrm{OH})_2 + \mathrm{H}^+ \label{react:feoh_feoh2}
\end{equation}
These equations are governed by their reaction rates: $k_{fe}$ (forward) and $k_{fe}'$ (backward) for Reaction \ref{react:fe_feoh}, and $k_{feoh}$ for Reaction \ref{react:feoh_feoh2}. Using these rate constants, the reaction rates can be written as \citep{Laycock2001}:
\begin{align}
    R_{\mathrm{Fe}^{2+}}&=-k_{fe}C_{\mathrm{Fe}^{2+}}+k_{fe}'C_{\mathrm{FeOH}^+}C_{\mathrm{H}^+} \\
    R_{\mathrm{FeOH}^+}&=k_{fe}C_{Fe^{2+}}-C_{\mathrm{FeOH}^+}(k_{feoh}+k_{fe}'C_{\mathrm{H}^+})\\
    R_{\mathrm{H}^+,fe}&=k_{fe}C_{\mathrm{Fe}^{2+}}-C_{\mathrm{FeOH}^+}(k_{fe}'C_{\mathrm{H}^+}-k_{feoh}) \label{eq:H_Part2}
\end{align}
From \cref{eq:water_react,eq:H_Part2}, a hydrogen reaction rate $R_{\mathrm{H}^+}$ can be defined that encompasses both water ionisation and corrosion contributions: $R_{\mathrm{H}^+} =  R_{\mathrm{H}^+,w} + R_{\mathrm{H}^+,fe}$. 

\subsection*{Metal sub-domain}
The deformation of the metal sub-domain is described by the balance of linear momentum; such that, neglecting body forces and inertia terms, 
\begin{equation}
    \bm{\nabla}\cdot\bm{\sigma} = \bm{0}
    \label{eq:mombalance}
\end{equation}
where the stress tensor $\bm{\sigma}$ is obtained assuming linear-elastic material behaviour. As it is typically the case, no direct influence of hydrogen on the elastic behaviour of the solid is defined. \\

For the description of the hydrogen present within the metal, we start with the chemical potential of the dissolved hydrogen:
\begin{equation}
    \mu = \mu_0 + RT \;\mathrm{ln}\left(\frac{\theta_L}{1-\theta_L}\right) - \overline{V}_H\sigma_H
    \label{eq:chem_pot}
\end{equation}
using the lattice site occupancy $\theta_L=C_L/N_L$, the hydrostatic stresses $\sigma_H=\mathrm{tr}(\bm{\sigma})/3$, the lattice sites density $N_L$, and the partial molar volume of hydrogen $\overline{V}_H$. This chemical potential is used to calculate the hydrogen fluxes through:
\begin{equation}
    \mathbf{J}_L = -\frac{D_L C_L}{RT}\bm{\nabla}\mu
    \label{eq:diff_from_potential}
\end{equation}
where $D_L$ is the lattice diffusivity. These definitions result in the following mass balance for the lattice hydrogen:
\begin{equation}
    \dot{C}_L + \bm{\nabla}\cdot\left(-\frac{D_L}{1-C_L/N_L} \bm{\nabla}C_L \right) + \bm{\nabla}\cdot\left(\frac{D_L C_L \overline{V}_H}{RT}\bm{\nabla}\sigma_H\right) = 0
    \label{eq:massbalance_lattice}
\end{equation}
In contrast to conventional formulations \citep{Turnbull1993, Turnbull2015, Diaz2016}, we do not assume a low lattice occupancy and as a result we obtain a hydrogen transport equation that depends not only on the concentration gradient but also on the concentration itself through the addition of a $1/(1-C_L/N_L)$ factor. This allows the present formulation to be valid over a larger range of parameters, increasing the local diffusivity to prevent the lattice hydrogen atoms to exceed the number of lattice sites (for instance, as caused by the presence of high hydrostatic stress gradients). A consequence of this is that the analytical expression for the hydrogen concentration at steady state is no longer estimated using $C_L = C_0\;\mathrm{exp}\left(\overline{V}_H\sigma_H/ (RT)\right)$ \citep{Kristensen2020a} (where $C_0$ is the concentration away from stress concentrators), but from setting \cref{eq:chem_pot} to be constant, resulting in:
\begin{equation}
    C_L=\frac{1-C_L/N_L}{1-C_0/N_L} C_0 \, \mathrm{exp}\left(\frac{\overline{V}_H\sigma_H}{RT} \right) 
    =  
    \frac{1}{1+\frac{C_0/N_L}{1-C_0/N_L}\mathrm{exp}\left(\frac{\overline{V}_H\sigma_H}{RT}\right)}\cdot C_0 \, \mathrm{exp}\left(\frac{\overline{V}_H\sigma_H}{RT} \right) 
    \label{eq:CMAX}
\end{equation}
While \cref{eq:CMAX} is not directly used within the model and thus solely provided for context, it shows that the maximum interstitial lattice concentration will tend closer to the reference concentration $C_0$ when these concentrations become closer to the number of interstitial lattice sites. As such, the influence of stress concentrators is less pronounced for higher $C_0$ values. It should also be noted that hydrogen sequestration in microstructural traps is not accounted for in the present model. While hydrogen traps can easily be included within \cref{eq:massbalance_lattice} \citep{Oriani1970, Turnbull2012, Fernandez-Sousa2020}, we have chosen not to do so to retain the focus on the stability and convergence behaviour of the electrolyte and the metal-electrolyte interface.

\subsection*{Metal-electrolyte interface}
At the metal-electrolyte interface, we consider the following reactions \citep{Elhamid2000, Danaee2011, Liu2014, Martinez-Paneda2020, Lasia1995}:
\begin{alignat}{2}
 \text{Volmer (acid):} && \mathrm{H}^+ + \mathrm{M} + \mathrm{e}^- &\xrightleftharpoons[k_{Va}']{k_{Va}} \mathrm{MH}_{ads} \label{react:1} \\
  \text{Heyrovsky (acid):} && \qquad \mathrm{H}^+ + \mathrm{e}^- + \mathrm{MH}_{ads}&\xrightleftharpoons[k_{Ha}']{k_{Ha}} \mathrm{M} + \mathrm{H}_2 \label{react:2} \\
    \text{Volmer (base):} &&  \mathrm{H}_2\mathrm{O} + \mathrm{M} + \mathrm{e}^- &\xrightleftharpoons[k_{Vb}']{k_{Vb}} \mathrm{MH}_{ads} + \mathrm{OH}^- \label{react:5} \\
   \text{Heyrovsky (base):} && \qquad  \mathrm{H}_2\mathrm{O} + \mathrm{e}^- + \mathrm{MH}_{ads}&\xrightleftharpoons[k_{Hb}']{k_{Hb}} \mathrm{M} + \mathrm{H}_2 + \mathrm{OH}^- \label{react:6} \\
    \text{Tafel:} && 2 \mathrm{MH}_{ads} &\xrightleftharpoons[k_T']{k_T} 2\mathrm{M} + \mathrm{H}_2 \label{react:3} \\
   \text{Absorption:} && \mathrm{MH}_{ads} &\xrightleftharpoons[k_A']{k_A} \mathrm{MH}_{abs}  \label{react:4} \\
   \text{Corrosion:} && \qquad  \mathrm{Fe}^{2+}+2\mathrm{e}^- &\xrightleftharpoons[k_c']{k_c} \mathrm{Fe} \label{react:7}
\end{alignat}
where the Volmer reactions are the main source of adsorbed hydrogen, either through the acidic Volmer reaction for low pH environments or through the much slower basic Volmer reaction for alkaline environments. For a low surface coverage and highly negative metal potentials, both Heyrovsky reactions are the main sink of adsorbed hydrogen, whereas for higher surface coverage the Tafel reaction becomes a more prominent hydrogen sink. On the other side, the absorption reaction characterises the quantity of hydrogen that goes from being attached to the surface to entering the bulk metal. For the corrosion reaction, we assume its rate to be insufficient to significantly dissolve the metal. While it is possible to include metal dissolution due to corrosion through smeared approaches such as phase field \citep{Cui2021, Shahmardi2021, Huang2022} or interface tracking schemes \citep{Duddu2011, Buoni2010}, we choose here to focus on the stability of the surface reactions when they are explicitly represented within the geometry. For the same reason, no passivation or protective layer development/dissolution is included within the context of the corrosion reactions. However, these phenomena can be readily included by adding or altering the corresponding electro-chemical reactions. \citep{Sohail2021, Ansari2019} Oxygen reduction reactions have also not been included. As is common within numerical simulations and experimental setups, \citep{Mai2016, Duddu2016, Malki2008} a constant electric potential is imposed on the metal. By applying this potential, the anodic and cathodic reaction rates decouple, and electric currents are allowed to enter and leave through the metal (the corrosion reaction can produce electrons, but these do not necessarily need to be consumed by the hydrogen and oxygen reactions). This reduces the role of the oxygen reduction reaction to providing a source of $\mathrm{OH}^-$ ions, which will locally increase the $\mathrm{pH}$ slightly but otherwise not influence the results. As such, the effect of not including the oxygen reduction reaction is expected to be limited.

The reaction rates for the reactions (\ref{react:1})-(\ref{react:7}) are given by \citep{Hageman2022}:
\begin{alignat}{4}
\nonumber && && & \qquad\mathrm{Forward} &&  \qquad\qquad \mathrm{Backward} \\
    \mathrm{Volmer (acid):} && \quad && \nu_{Va} &= k_{Va} C_{\mathrm{H}^+}(1-\theta_{ads})e^{-\alpha_{Va} \frac{\eta F}{RT}}\qquad
    && \nu_{Va}' = k_{Va}' \theta_{ads}e^{(1-\alpha_{Va}) \frac{\eta F}{RT}} \label{eq:react1}\\
    \mathrm{Heyrovsky (acid):} && && \nu_{Ha} &= k_{Ha} C_{\mathrm{H}^+}\theta_{ads}e^{-\alpha_{Ha} \frac{\eta F}{RT}}\qquad
    && \nu_{Ha}' = k_{Ha}' (1-\theta_{ads}) p_{\mathrm{H}_2} e^{(1-\alpha_{Ha}) \frac{\eta F}{RT}} \label{eq:react2}\\
    \mathrm{Volmer (base):} && && \nu_{Vb} &= k_{Vb} (1-\theta_{ads})e^{-\alpha_{Vb} \frac{\eta F}{RT}}\qquad
    && \nu_{Vb}' = k_{Vb}' C_{\mathrm{OH}^-} \theta_{ads}e^{(1-\alpha_{Vb}) \frac{\eta F}{RT}} \label{eq:react5}\\
    \mathrm{Heyrovsky (base):} && && \nu_{Hb} &= k_{Hb} \theta_{ads}e^{-\alpha_{Hb} \frac{\eta F}{RT}}\qquad
    && \nu_{Hb}' = k_{Hb}' (1-\theta_{ads}) p_{\mathrm{H}_2} C_{\mathrm{OH}^-} e^{(1-\alpha_{Hb}) \frac{\eta F}{RT}}  \label{eq:react6}\\
    \mathrm{Tafel:} && && \nu_T &= k_T\left|\theta_{ads}\right|\theta_{ads}\qquad
    && \nu_T' = k_T' (1-\theta_{ads})p_{\mathrm{H}_2} \label{eq:react3}\\
    \mathrm{Absorption:} && && \nu_A &= k_A (N_L - C_L)\theta_{ads}\qquad
    && \nu_A' = k_A' C_L (1-\theta_{ads}) \label{eq:react4}\\
    \mathrm{Corrosion:} && && \nu_{c} &= k_{c} C_{\mathrm{Fe}^{2+}}e^{-\alpha_{c} \frac{\eta F}{RT}} \qquad && \nu_{c}' = k_{c}' e^{(1-\alpha_{c}) \frac{\eta F}{RT}}   \label{eq:react7}
\end{alignat}
These reactions use constant forward and backward rate constants $k$ and $k'$, charge transfer coefficient $\alpha$, and equilibrium potential $E_{eq}$. The reaction rates depend on the ionic concentrations within the electrolyte, $C_\pi$, the hydrogen surface coverage, $\theta_{ads}$, the lattice hydrogen concentration $C_L$, and the metal and electrolyte potentials through the overpotential, $\eta=E_m-\varphi-E_{eq}$. As a result, they depend on and provide the coupling between the electrolyte and metal domain through the reaction fluxes as: 
\begin{align}
    \overline{J}_{\mathrm{H}^+} &= -(\nu_{Va} - \nu_{Va}') - \nu_{Ha}  \\
    \overline{J}_{\mathrm{OH}^-} &= \nu_{Vb} - \nu_{Vb}' + \nu_{Hb}   \\
    \overline{J}_{\mathrm{Fe}^{2+}} &= \nu_{c}'-\nu_c\\
    \overline{J}_L &= \nu_A - \nu_A'
\end{align}
where we have neglected the backwards reaction rates of the Heyrovsky and Tafel reactions based on the assumption that the hydrogen gas produced disappears from the domain, resulting in a negligible reaction rate ($\nu_T'=\nu_{Ha}'=\nu_{Hb}'=0$). 
Finally, the surface mass balance for the adsorbed hydrogen reads:
\begin{equation}
    N_{ads} \dot{\theta}_{ads} - (\nu_{Va}-\nu_{Va}') + \nu_{Ha} + 2 \nu_T + (\nu_A-\nu_A') - (\nu_{Vb}-\nu_{Vb}') + \nu_{Hb} = 0
    \label{eq:massbalanceinterface}
\end{equation}

\subsection*{Discretisation}
\label{sec:disc}
We discretise the governing equations presented in the previous sub-sections using the finite element method, interpolating the electrolyte potential and concentrations using quadratic triangular elements as:
\begin{equation}
    \varphi = \sum_{el} \mathbf{N}_\varphi^{el} \bm{\upvarphi}^{el} \qquad C_\pi = \sum_{el} \mathbf{N}_c^{el} \mathbf{C}_\pi
\end{equation}
Similarly, for the metal sub-domain, the displacements and lattice hydrogen concentration are interpolated as:
\begin{equation}
    \mathbf{u} = \sum_{el} \bm{N}_u^{el} \mathbf{u}^{el} \qquad C_L = \sum_{el} \mathbf{N}_L^{el} \mathbf{C}_L^{el}
\end{equation}
and we represent the hydrogen surface coverage using one-dimensional quadratic line elements at the metal-electrolyte interface, such that
\begin{equation}
    \theta_{ads} = \sum_{iel} \mathbf{N}_\theta^{el} \bm{\uptheta}^{el}
\end{equation}
In addition to this spatial discretisation, we perform the temporal discretisation using a backward Euler scheme, evaluating the governing equations at $t+\Delta t$ and using the time derivative:
\begin{equation}
    \dot{\square} = \frac{\square^{t+\Delta t}-\square^t}{\Delta t}
\end{equation}

Using these discretisations, the governing equations for the metal domain, Eqs. (\ref{eq:mombalance}) and (\ref{eq:massbalance_lattice}), are transformed into their discretised weak form, with the following force components
\begin{equation}
    \mathbf{f}_u = \int_{\Omega_m} \bm{B}_u^T\bm{D}\bm{B}_u\mathbf{u}^{t+\Delta t}\;\mathrm{d}\Omega_m - \int_{\Gamma_m}\bm{N}_u^T \mathbf{t}_{ext} \; \mathrm{d}\Gamma_m = \bm{0} \label{eq:f_u}
\end{equation}
\begin{equation}
\begin{split}
    \mathbf{f}_L = &\int_{\Omega_m} \frac{1}{\Delta t}\mathbf{N}_L^T\mathbf{N}_L\left(\mathbf{C}_L^{t+\Delta t}-\mathbf{C}_L^t\right)\;\mathrm{d}\Omega_m + \int_{\Omega_m} \frac{D_L}{1-\mathbf{N}_L \mathbf{C}_L^{t+\Delta t}/N_L}\left(\bm{\nabla}\mathbf{N}_L\right)^T\bm{\nabla}\mathbf{N}_L\mathbf{C}_L^{t+\Delta t}\;\mathrm{d}\Omega_{m}\\ &- \int_{\Omega_m} \frac{E}{3(1-2\nu)}\frac{D_L \overline{V}_H}{RT} \left(\bm{\nabla}\mathbf{N}_L\right)^T \left(\mathbf{N}_L \mathbf{C}_L^{t+\Delta t}\right)\bm{B}_u^* \mathbf{u}^{t+\Delta t} \; \mathrm{d}\Omega_m  \\
    &- \int_{\Gamma_{int}} \mathbf{N}_L^T \left( \nu_A^{t+\Delta t} - \nu_A'^{t+\Delta t} \right) \; \mathrm{d}\Gamma_{int} - {\int_{\Gamma_m}} \mathbf{N}_L^T j_L \; \mathrm{d}\Gamma_m = \mathbf{0}
    \end{split} \label{eq:f_L}
\end{equation}
where the boundary conditions on the external boundary $\Gamma_m$ are defined as the external tractions $\mathbf{t}_{ext}$ and external hydrogen influx $j_L$. The displacement to strain and displacement to strain gradient mapping matrices are defined under plane-strain conditions as:
\begin{align}
    \bm{B}_u &= \begin{bmatrix} 
    \frac{\partial N_{u1}}{\partial x} & \frac{\partial N_{u2}}{\partial x} 
    & \cdot\cdot\cdot & 
    0 & 0 
    & \cdot\cdot\cdot\\ 
    0 & 0 & \cdot\cdot\cdot &
    \frac{\partial N_{u1}}{\partial y} & \frac{\partial N_{u2}}{\partial y} & \cdot\cdot\cdot \\ 0 & 0 & \cdot\cdot\cdot & 0 & 0 & \cdot\cdot\cdot \\ \frac{\partial N_{u1}}{\partial y} & \frac{\partial N_{u2}}{\partial y} & \cdot\cdot\cdot & \frac{\partial N_{u1}}{\partial x}& \frac{\partial N_{u2}}{\partial x} & \cdot\cdot\cdot \end{bmatrix} \\
    \bm{B}_u^* &= \begin{bmatrix} \frac{\partial^2 N_{u1}}{\partial x^2} & \frac{\partial^2 N_{u2}}{\partial x^2} & \cdot\cdot\cdot & \frac{\partial^2 N_{u1}}{\partial x\partial y} & \frac{\partial^2 N_{u2}}{\partial x\partial y} & \cdot \cdot \cdot \\
    \frac{\partial^2 N_{u1}}{\partial x\partial y} & \frac{\partial^2 N_{u2}}{\partial x \partial y} & \cdot\cdot\cdot & \frac{\partial^2 N_{u1}}{\partial y^2} & \frac{\partial^2 N_{u2}}{\partial y^2} & \cdot \cdot \cdot
    \end{bmatrix}
\end{align}
such that the strains are given by $\bm{\varepsilon}=\bm{B}_u [\mathbf{u}_x; \; \mathbf{u}_y]$ and the hydrostatic stress gradient as $\bm{\nabla}\sigma_H=E/(3(1-2\nu))\bm{B}_u^*[\mathbf{u}_x; \; \mathbf{u}_y]$.\\

For the electrolyte, the discretised weak forms of the mass balances,  \cref{eq:nernstplanck}, and electroneutrality condition,  \cref{eq:electroneutrality}, are given by:
\begin{equation} \begin{split}
    \mathbf{f}_{c\pi} = &\int_{\Omega_{e}} \frac{1}{\Delta t}\mathbf{N}_C^T \mathbf{N}_C \left(\mathbf{C}_\pi^{t+\Delta t} - \mathbf{C}_\pi^t\right)\;\mathrm{d}\Omega_{e} + \int_{\Omega_{e}} D_\pi \left(\bm{\nabla}\mathbf{N}_c\right)^T\bm{\nabla}\mathbf{N}_C\mathbf{C}_\pi^{t+\Delta t} \; \mathrm{d}\Omega_{e} \\
    &+ \int_{\Omega_{e}} \frac{D_\pi z_\pi F}{RT} \left(\bm{\nabla}\mathbf{N}_c\right)^T \left(\mathbf{N}_C \mathbf{C}_\pi^{t+\Delta t}\right) \bm{\nabla}\mathbf{N}_\varphi \bm{\upvarphi}^{t+\Delta t}\;\mathrm{d}\Omega_{e} + \mathbf{R}_{\pi} + \mathbf{\upnu}_{\pi} - \int_{\Gamma_{e}} \mathbf{N}_c j_c \; \mathrm{d}\Gamma_{e} = \mathbf{0}
\end{split} \label{eq:f_c} \end{equation} 
\begin{equation}
    \bm{f}_\varphi = \int_{\Omega_{e}} \sum_\pi z_\pi  \mathbf{N}_\varphi^T\mathbf{N}_c\mathbf{C}_\pi^{t+\Delta t} \; \mathrm{d}\Omega_{e} = \mathbf{0} \label{eq:f_phi}
\end{equation}
where the bulk reaction rates are given by:
\begin{subequations}
\begin{align}
\begin{split}
    \mathbf{R}_{\mathrm{H}^+} &= -\int_{\Omega_{e}} k_{eq}\mathbf{N}_c^T \left(K_w-\left(\mathbf{N}_c\mathbf{C}_{\mathrm{H}^+}^{t+\Delta t}\right)\left(\mathbf{N}_c\mathbf{C}_{\mathrm{OH}^-}^{t+\Delta t}\right)\right)\;\mathrm{d}\Omega_{e} \\&\quad- \int_{\Omega_{e}} \mathbf{N}_c^T \left( k_{fe}\mathbf{N}_c\mathbf{C}_{\mathrm{Fe}^{2+}}^{t+\Delta t}-\mathbf{N}_c\mathbf{C}_{\mathrm{FeOH}^+}^{t+\Delta t}\left(k'_{fe}\mathbf{N}_c\mathbf{C}_{\mathrm{H}^+}^{t+\Delta t} - k_{feoh}\right) \right)\;\mathrm{d}\Omega_{e}\end{split} \label{eq:RH} \\
    \mathbf{R}_{\mathrm{OH}^-} &= -\int_{\Omega_{e}} \mathbf{N}_c^T k_{eq} \left(K_w-\left(\mathbf{N}_c\mathbf{C}_{\mathrm{H}^+}^{t+\Delta t}\right)\left(\mathbf{N}_c\mathbf{C}_{\mathrm{OH}^-}^{t+\Delta t}\right)\right)\;\mathrm{d}\Omega_{e} \label{eq:ROH} \\
    \mathbf{R}_{\mathrm{Fe}^{2+}} &= \int_{\Omega_{e}} \mathbf{N}_c^T \left(k_{fe} \mathbf{N}_c \mathbf{C}_{\mathrm{Fe}^{2+}} - k_{fe}' \mathbf{N}_c \mathbf{C}_{\mathrm{FeOH}^+}^{t+\Delta t}\mathbf{N}_C\mathbf{C}_{\mathrm{H}^+}^{t+\Delta t}\right) \; \mathrm{d}\Omega_{e} \label{eq:RFE} \\
    \mathbf{R}_{\mathrm{FeOH}^+} &= - \int_{\Omega_{e}} \mathbf{N}_c^T \left( k_{fe}\mathbf{N}_c\mathbf{C}_{\mathrm{Fe}^{2+}}^{t+\Delta t}-\mathbf{N}_c\mathbf{C}_{\mathrm{FeOH}^+}^{t+\Delta t}\left(k'_{fe}\mathbf{N}_c\mathbf{C}_{\mathrm{H}^+}^{t+\Delta t} + k_{feoh}\right) \right)\;\mathrm{d}\Omega_{e} \label{eq:RFEOH} \\
    \mathbf{R}_{\mathrm{Na}^+} &= \mathbf{R}_{\mathrm{Cl}^-} = \mathbf{0} \label{eq:RNACL}
\end{align}
\end{subequations}
and the surface reaction terms by:
\begin{subequations}
\begin{align}
    \mathbf{\upnu}_{\mathrm{H}^+}    &=  \int_{\Gamma_{int}} \mathbf{N}_C^T \left(\nu_{Va}^{t+\Delta t}-{\nu'}_{Va}^{t+\Delta t}+\nu_{Ha}^{t+\Delta t}\right)\;\mathrm{d}\Gamma_{int}\\
    \mathbf{\upnu}_{\mathrm{OH}^-}   &=  -\int_{\Gamma_{int}} \mathbf{N}_C^T \left(\nu_{Vb}^{t+\Delta t}-{\nu'}_{Vb}^{t+\Delta t}+\nu_{Hb}^{t+\Delta t}\right)\;\mathrm{d}\Gamma_{int}\\
    \mathbf{\upnu}_{\mathrm{Fe}^{2+}}&= \int_{\Gamma_{int}} \mathbf{N}_C^T \left(\nu_{c}^{t+\Delta t}-{\nu'}_{c}^{t+\Delta t}\right)\;\mathrm{d}\Gamma_{int}\\
    \mathbf{\upnu}_{\mathrm{Na}^+}   &=\nu_{\mathrm{Cl}^-}=\nu_{\mathrm{FeOH}^+} = \mathbf{0}
\end{align}
\end{subequations}

Finally, we have the interfacial mass balance for the adsorbed hydrogen, which is given in discretised form as:
\begin{equation}
\begin{split}
    \mathbf{f}_{\theta}=&\int_{\Gamma_{int}} \frac{N_{ads}}{\Delta t}\mathbf{N}_\theta^T\mathbf{N}_\theta \left(\bm{\uptheta}^{t+\Delta t}-\bm{\uptheta}^t\right)\;\mathrm{d}\Gamma_{int} \\- &\int_{\Gamma_{int}} \mathbf{N}_\theta^T \left((\nu_{Va}-\nu_{Va}') - \nu_{Ha} - 2\nu_T - (\nu_A-\nu_A') + (\nu_{Vb}-\nu_{Vb}') - \nu_{Hb}\right)\; \mathrm{d}\Gamma_{int}=\bm{0}
    \label{eq:f_theta} \end{split} 
\end{equation}

The mass balances from \cref{eq:f_L,eq:f_c,eq:f_theta}, the momentum balance from \cref{eq:f_u}, and the charge balance from \cref{eq:f_phi} together fully describe the behaviour of the electrolyte, metal and interface. To eliminate any stability issues originating from the coupling between the domains, both domains are solved at the same time in a monolithic manner. To solve the resulting system of equations, an iterative Newton-Raphson scheme is used. Details of the solution scheme are given in \cref{sec:matrices}, together with the formulation of the tangential matrices employed.

\section*{Integration schemes}
\label{sec:integration}

\begin{algorithm}[t]
\caption{Overview of Gauss and lumped integration schemes}\label{alg:integration_scheme}
\begin{algorithmic}[1]
\State Start of element assembly
\State Set element flux vector and lumped weight vector to 0: $\mathbf{q}=\mathbf{W}=\mathbf{0}$
\For{integration point $ip=1:n_{ip}$}
\State Calculate Gauss weights $w_{ip}$, and shape functions $\bm{N}_{ip}$
\State Calculate reaction rates $\nu_{ip}$ based on local integration point values ($C_{\mathrm{H}^+}=\mathbf{N}_{ip}\mathbf{C}_{\mathrm{H}^+}$, etc.)
\State Perform non-lumped integrations: $\mathbf{q}=\mathbf{q}+w_{ip}\mathbf{N}^T\nu_{ip}$
\State Calculate lumped weights: $\mathbf{W}=\mathbf{W} + w_{ip}\mathbf{N}_{ip}$
\EndFor
\For{node $nd=1:n_{nodes}$}
\State Calculate reaction rates $\nu_{nd}$ based on nodal values (using $\mathbf{C}_{\mathrm{H}^+}(nd)$, etc.)
\State Perform lumped integrations: $\mathbf{q}(nd)=\mathbf{q}(nd)+\mathbf{W}(nd)\;\nu_{nd}$
\EndFor
\State Go to next element
\end{algorithmic}
\end{algorithm}

In the context of the finite element method, the weak form equations are typically integrated using a Gauss integration scheme; evaluating the value of the functions at set points within the element and using these values to estimate the integral. However, as discussed below, this results in spurious oscillations throughout the electrolyte. These oscillations originate from two main sources. One is the water auto-ionisation reaction - the first term in \cref{eq:RH} and \cref{eq:ROH}. More specifically, oscillations are a consequence of the high value that must be assigned to the penalty reaction rate term $k_{eq}$ to ensure equilibrium. A second source is the absorption reaction, as a result of the high magnitudes of the forward and backward reaction rates, $k_A$ and $k_A'$. These are relevant for the solution of the lattice hydrogen concentration $C_L$, see \cref{eq:f_L}, and the surface coverage, see \cref{eq:f_theta}, bringing instabilities to the coupling between these two fields.\\

To prevent these oscillations, we use a lumped integration scheme \citep{Schellekens1993, Vignollet2015}. This scheme first constructs the lumped weight vector by employing a standard Gauss integration scheme on the element under consideration as:
\begin{equation}
    \mathbf{W} = \int_{\Omega_{el}} \mathbf{N}^T\;\mathrm{d}\Omega_{el} = \sum_{ip} w_{ip} \mathbf{N}^T(\mathbf{x}_{ip})
\end{equation}
where $w_{ip}$ indicates the weighting factor of the integration point, and $\mathbf{x}_{ip}$ are the coordinates of the current integration point. The nodal integration weights are then used to integrate the reactions based on the values in the nodes. For example, the reaction rate of $\mathrm{OH}^-$, \cref{eq:ROH}, is reformulated as:
\begin{equation}
    \bm{R}_{OH^-} = -\int_{\Omega_{e}} \mathbf{N}_c^T k_{eq} \left(K_w-\left(\mathbf{N}_c\mathbf{C}_{\mathrm{H}^+}^{t+\Delta t}\right)\left(\mathbf{N}_c\mathbf{C}_{\mathrm{OH}^-}^{t+\Delta t}\right)\right)\;\mathrm{d}\Omega_{e} = -\sum_{el} \sum_{nd} \mathbf{W}(nd) k_{eq} \left(K_W - \mathbf{C}_{\mathrm{H}^+}(nd) \mathbf{C}_{\mathrm{OH}^-}(nd) \right)
\end{equation}
and similarly for the water auto-ionisation term, \cref{eq:RH}. Here, $\mathbf{\square}(nd)$ is used to indicate the nodal component of vector $\mathbf{\square}$, and the resulting magnitude is allocated to the force vector index associated with the $\mathrm{OH}^-$ degree of freedom at node $nd$. This force vector assembly process is summarised in \cref{alg:integration_scheme} and the \texttt{MATLAB} finite element code developed is openly shared to facilitate understanding and uptake\footnote{The code is available to download at \url{www.imperial.ac.uk/mechanics-materials/codes} and \url{www.empaneda.com/codes}.}.\\

Applying a similar scheme to the absorption reaction terms in \cref{eq:f_L,eq:f_theta} leads to the following sum over all the nodes in the elements at the interface:
\begin{equation}
    \int_{\Gamma_{int}} \mathbf{N}_x^T (\nu_{Va}-\nu_{Va}')\;\mathrm{d}\Gamma_{int} = \sum_{iel} \sum_{nd} \bm{W}(nd) \Big( k_A (N_L - \mathbf{C}_L(nd)\;)\bm{\uptheta}_{ads}(nd) - k_A' \mathbf{C}_L(nd) (1-\bm{\uptheta}_{ads}(nd)\;)\Big)
    \label{eq:AbsLI}
\end{equation}
where $\mathbf{N}_x$ represents the interpolation functions $\mathbf{N}_\theta$ for the terms associated with the surface coverage, and $\mathbf{N}_L$ for the terms corresponding to the interstitial lattice hydrogen concentration. As a result, all off-diagonal terms associated with these reactions are taken out from the stiffness matrix. In this way, the interactions between neighbouring nodes that take place through these high reaction rates are eliminated.\\

\begin{figure}
    \centering
    \begin{subfigure}{7.5cm}
         \centering
         \textbf{Gauss integration}
         \includegraphics[width=7cm]{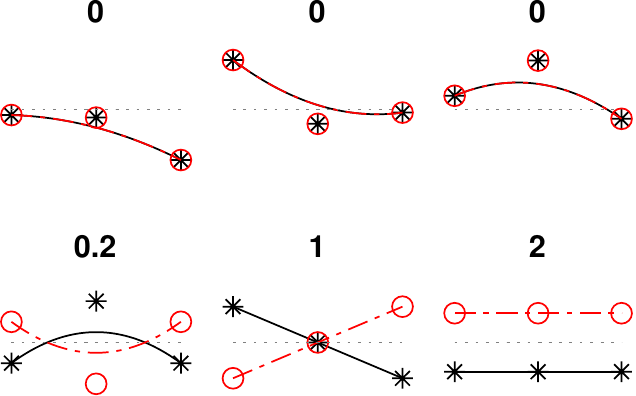}
         \caption{}
         \label{fig:Eigen_NoMass_Gauss}
    \end{subfigure}
    \begin{subfigure}{7.5cm}
         \centering
         \textbf{Lumped integration}
         \includegraphics[width=7cm]{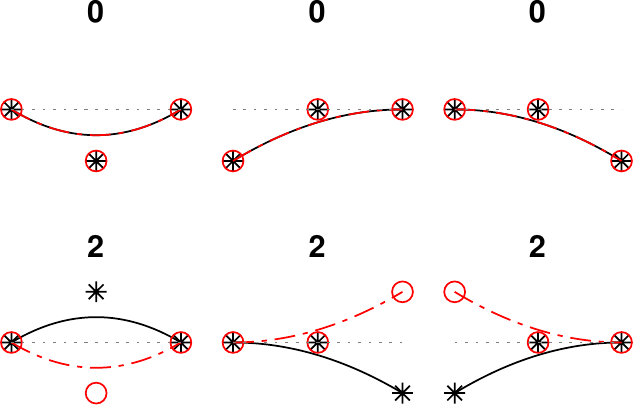}
         \caption{}
         \label{fig:Eigen_NoMass_Lumped}
     \end{subfigure}
    \caption{Eigenvalues and modes obtained for the matrices from \cref{eq:KMats} using: (a) a Gauss integration scheme, and (b) the lumped integration scheme. Red dashed lines are used for the changes in lattice hydrogen concentration, while black lines are used for the changes in surface occupancy relative to the zero concentration and occupancy (grey dotted lines). Red circles and black stars indicate the values within the nodes.}
    \label{fig:Eigen_NoMass}
\end{figure}

Let us proceed to illustrate the effect of the lumped integration scheme proposed by considering a simple example where only the hydrogen absorption reaction is modelled over a single element. In such scenario, the weak forms read: 
\begin{align}
    q_\theta & =  \int \mathbf{N}^T \left(k_A \left(N_L-\mathbf{N}_L \mathbf{C}_L\right) \mathbf{N}_\theta \mathbf{\uptheta} - k_A' \mathbf{N}_L\mathbf{C}_L \left(1-\mathbf{N}_{\theta}\mathbf{\uptheta}\right)\right) \; \mathrm{d}\Gamma = \mathbf{0} \\
    q_L & =  - \int \mathbf{N}^T \left(k_A \left(N_L-\mathbf{N}_L \mathbf{C}_L\right) \mathbf{N}_\theta \mathbf{\uptheta} - k_A' \mathbf{N}_L\mathbf{C}_L \left(1-\mathbf{N}_{\theta}\mathbf{\uptheta}\right)\right) \; \mathrm{d}\Gamma = \mathbf{0}
\end{align}
\noindent where $\mathbf{q}$ is used to denote the weak forms solely related to the reaction fluxes. We shall now calculate the tangential matrix terms for this system using Gauss and lumped integration. To this end, a single quadratic element is used, the solution vectors are defined as $\mathbf{C}_L=\mathbf{\uptheta}=\mathbf{0}$, and we set $k_A=k_A'=3$ and $N_L=1$, rendering:
\begin{equation}
    \bm{K}_{Gauss} = \begin{bmatrix}
        0.6 & 0.3 & 0.1 & -0.6 & -0.3 & -0.1\\
        0.3 & 0.4 & 0.3 & -0.3 & -0.4 & -0.3\\
        0.1 & 0.3 & 0.6 & -0.1 & -0.3 & -0.6\\
        -0.6 & -0.3 & -0.1 & 0.6 & 0.3 & 0.1\\
        -0.3 & -0.4 & -0.3 & 0.3 & 0.4 & 0.3\\
        -0.1 & -0.3 & -0.6 & 0.1 & 0.3 & 0.6
    \end{bmatrix} \quad     \bm{K}_{Lumped} = \begin{bmatrix}
        1 & 0 & 0 & -1 & 0 & 0\\
        0 & 1 & 0 & 0 & -1 & 0\\
        0 & 0 & 1 & 0 & 0 & -1\\
        -1 & 0 & 0 & 1 & 0 & 0\\
        0 & -1 & 0 & 0 & 1 & 0\\
        0 & 0 & -1 & 0 & 0 & 1
    \end{bmatrix} \label{eq:KMats}
\end{equation}
It can be seen that both integration schemes result in a similar hydrogen flux (e.g., for each row, the same magnitude is obtained when adding the terms in columns 1-3). As expected, the lumped scheme results in a tangential matrix with all terms concentrated on the diagonals of each degree-of-freedom sub-matrix. This is in sharp contrast with the Gauss scheme, which results in a dense matrix. Differences are also showcased in \cref{fig:Eigen_NoMass}, which shows the eigenvalues and eigenmodes obtained from the matrices in \cref{eq:KMats}. These eigenmodes indicate the manner in which concentrations are expected to change, and the eigenvalues provide an indication of the frequency response of these changes. Since capacity terms are not included, three ``free body motion'' eigenvalues are present, representing the modes in which both the lattice and adsorbed hydrogen concentrations increase equally. Looking at the non-zero eigenvalues and accompanying eigenmodes, one can see clear differences between the lumped and Gauss integration schemes. The lumped scheme obtains three equal eigenvalues, corresponding to a single set of nodes transferring hydrogen from the surface to the metal lattice. As all eigenmodes are equal, this system response will allow for an oscillation-free transfer of hydrogen from the metal surface into the interstitial lattice sites. In contrast, Gaussian integration results in three different eigenvalues. The lower two of these will produce oscillations due to the way the three nodes are coupled. Furthermore, they do not transfer hydrogen from the surface to the lattice, but instead redistribute it between neighbouring nodes. Only the inclusion of the highest eigenmode allows for the transfer of hydrogen. These low eigenvalues corresponding to oscillatory results indicate that the hydrogen absorption reaction is prone to spurious oscillations when this term becomes dominant; this is not observed for the lumped scheme.

\FloatBarrier
\section*{On the role of lumped integration in simulating hydrogen absorption}
\label{sec:ParametricSweep}

\begin{table}%[p!]
 \caption{Material and ionic transport parameters used in all cases reported within this paper.}
\label{table:params}
        \centering
\begin{tabular}{ |l l|l|  }
 \hline 
  Parameter & & Value\\
 \hline  \hline
 Young's Modulus & $E$ & $200\;\mathrm{GPa}$\\
 Poisson ratio & $\nu$ & $0.3$\\
 \hline
 $\mathrm{H}^+$ diffusion coefficient & $D_{\mathrm{H}^+}$ & $9.3\cdot10^{-9}\;\mathrm{m}^2/\mathrm{s}$ \\
 $\mathrm{OH}^-$ diffusion coefficient & $D_{\mathrm{OH}^-}$ & $5.3\cdot10^{-9}\;\mathrm{m}^2/\mathrm{s}$ \\
 $\mathrm{Na}^+$ diffusion coefficient & $D_{\mathrm{Na}^+}$ & $1.3\cdot10^{-9}\;\mathrm{m}^2/\mathrm{s}$ \\
 $\mathrm{Cl}^-$ diffusion coefficient & $D_{\mathrm{Cl}^-}$ & $2\cdot10^{-9}\;\mathrm{m}^2/\mathrm{s}$ \\
$\mathrm{Fe}^{2+}$ diffusion coefficient & $D_{\mathrm{Fe}^{2+}}$ & $1.4\cdot10^{-9}\;\mathrm{m}^2/\mathrm{s}$ \\
$\mathrm{FeOH}^+$ diffusion coefficient & $D_{\mathrm{FeOH}^+}$ & $10^{-9}\;\mathrm{m}^2/\mathrm{s}$ \\
\hline
Partial molar volume & $\overline{V}_H$ & $2\cdot10^{-6}\;\mathrm{mol}/\mathrm{m}^3$\\
Surface adsorption sites & $N_{ads}$ & $10^{-3}\;\mathrm{mol}/\mathrm{m}^2$ \\
Lattice sites & $N_L$ & $10^6\;\mathrm{mol}/\mathrm{m}^3$ \\
Lattice diffusion coefficient & $D_L$ & $10^{-9}\;\mathrm{m}^2/\mathrm{s}$ \\
Temperature & $T$ & $293.15\;\mathrm{K}$\\
 \hline
\end{tabular}
\end{table}

\begin{table}
%\centering
 \caption{Reaction rate constants used throughout the paper (backward $k$ and forward $k'$. For the cases in which a parametric sweep is performed, all reaction constants are taken from this table except those explicitly stated to be different.} \label{tab:reactionsused}
        \centering
\begin{tabular}{ |l|l|l|l|l|  }
 \hline
  Reaction & $k$ & $k'$ & $\alpha$ & $E_{eq}$\\
  \hline \hline 
 $\nu_{Va}$ & $1\cdot10^{-4}\; \mathrm{m}/\mathrm{s}$ & $1\cdot10^{-10}\;\mathrm{mol/(m}^2\mathrm{s)}$ & $0.5$ & $0\;\mathrm{V}_{\mathrm{SHE}}$ \\
 %\Xhline{0.1pt}  
 $\nu_{Ha}$ & $1\cdot10^{-10} \;\mathrm{m/s}\;\;$ & $0 \;\mathrm{mol/(m}^2\mathrm{Pa\; s)}$ & $0.3$ & $0\;\mathrm{V}_{\mathrm{SHE}}$\\
 %\Xhline{0.1pt}  
 $\nu_T$ & $1\cdot10^{-6} \;\mathrm{mol/(m}^2\mathrm{s)}$ & $0 \;\mathrm{mol/(m}^2\mathrm{s \;Pa}^{1/2})$ & $-$ & $-$ \\
  %\Xhline{0.1pt}  
 $\nu_A$ & $1\cdot10^3 \;\mathrm{m/s}$ & $7\cdot10^7 \;\mathrm{m/s}$ & $-$ & $-$ \\ 
  %\Xhline{0.1pt}  
 $\nu_{Vb}$ & $1\cdot10^{-8} \; \mathrm{mol/(m}^2\mathrm{s})$ & $1\cdot10^{-13} \;\mathrm{m/s}$ & $0.5$ & $0\;\mathrm{V}_{\mathrm{SHE}}$ \\
  %\Xhline{0.1pt}  
 $\nu_{Hb}$ & $1\cdot10^{-10} \;\mathrm{mol/(m}^2\mathrm{s)}$ & $0 \;\mathrm{m/(Pa \;s)}$ & $0.3$ & $0\;\mathrm{V}_{\mathrm{SHE}}$ \\
  %\Xhline{0.1pt}  
 $\nu_{Fe}$ & $1.5\cdot10^{-10}\;\mathrm{mol}/\mathrm{(m}^2\mathrm{s)}$ & $1.5\cdot10^{-10}\;\mathrm{m}/\mathrm{s}$ & $0.5$ & $-0.4\;\mathrm{V}_{\mathrm{SHE}}$ \\
 %\hline
 \hline
$k_{fe}$ & $0.1\;\mathrm{s}$ & $10^{-3}\;\mathrm{m}^3/(\mathrm{mol}\;\mathrm{s})$ &  &  \\
$k_{feoh}$ & $10^{-3} \;\mathrm{s}^{-1}$ & & &  \\
$k_{eq}$ & $10^6\;\mathrm{m}^3/(\mathrm{mol}\;\mathrm{s})$ & & & \\
\hline
\end{tabular}
\end{table}

\begin{figure}
    \centering
    \begin{subfigure}{7.5cm}
         \centering
         \textbf{Gauss}
         \includegraphics[width=7.5cm]{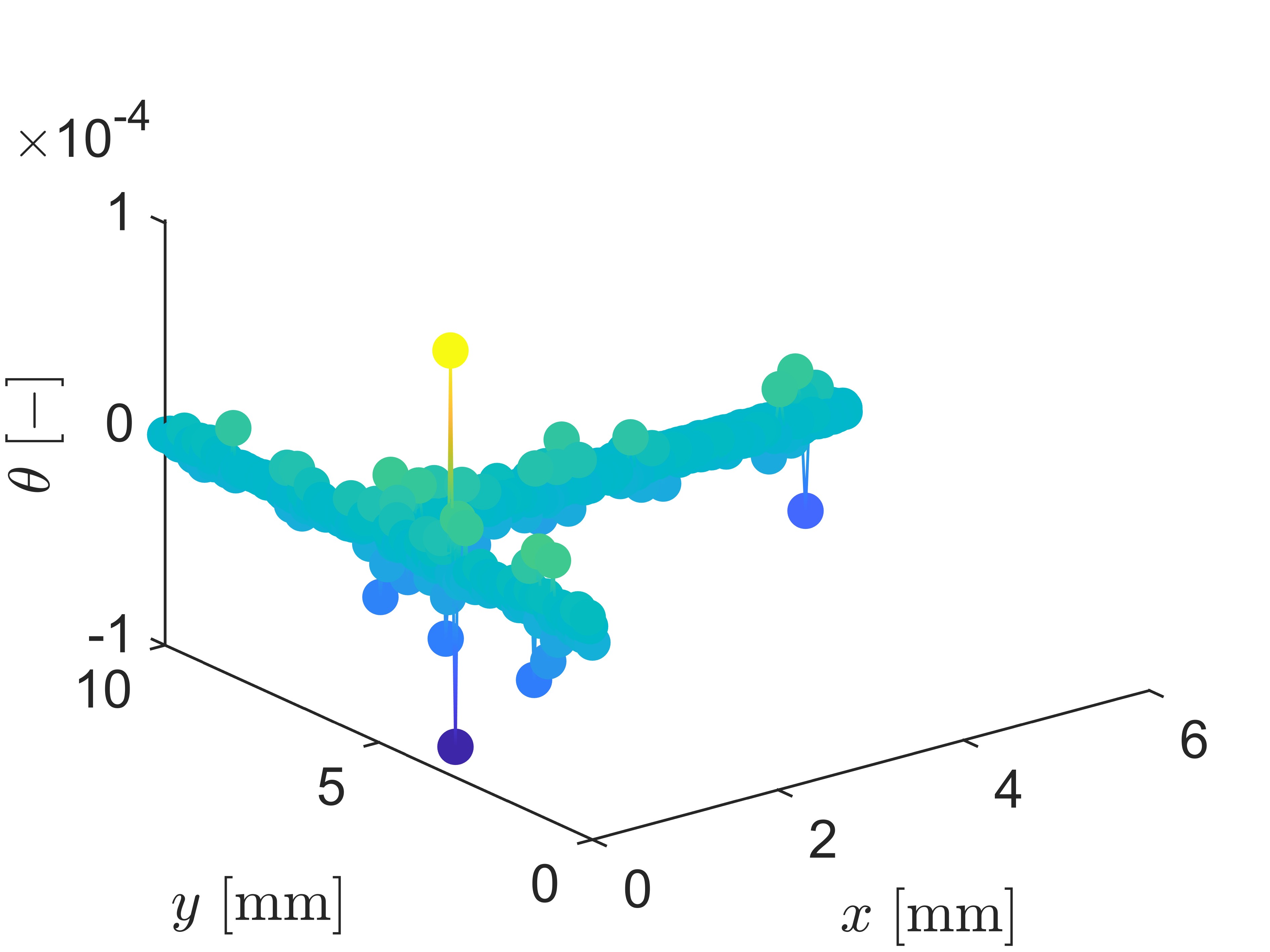}
         \caption{}
         \label{fig:effect_lumped_theta_Standard}
    \end{subfigure}
    \begin{subfigure}{7.5cm}
         \centering
         \textbf{Lumped absorption}
         \includegraphics[width=7.5cm]{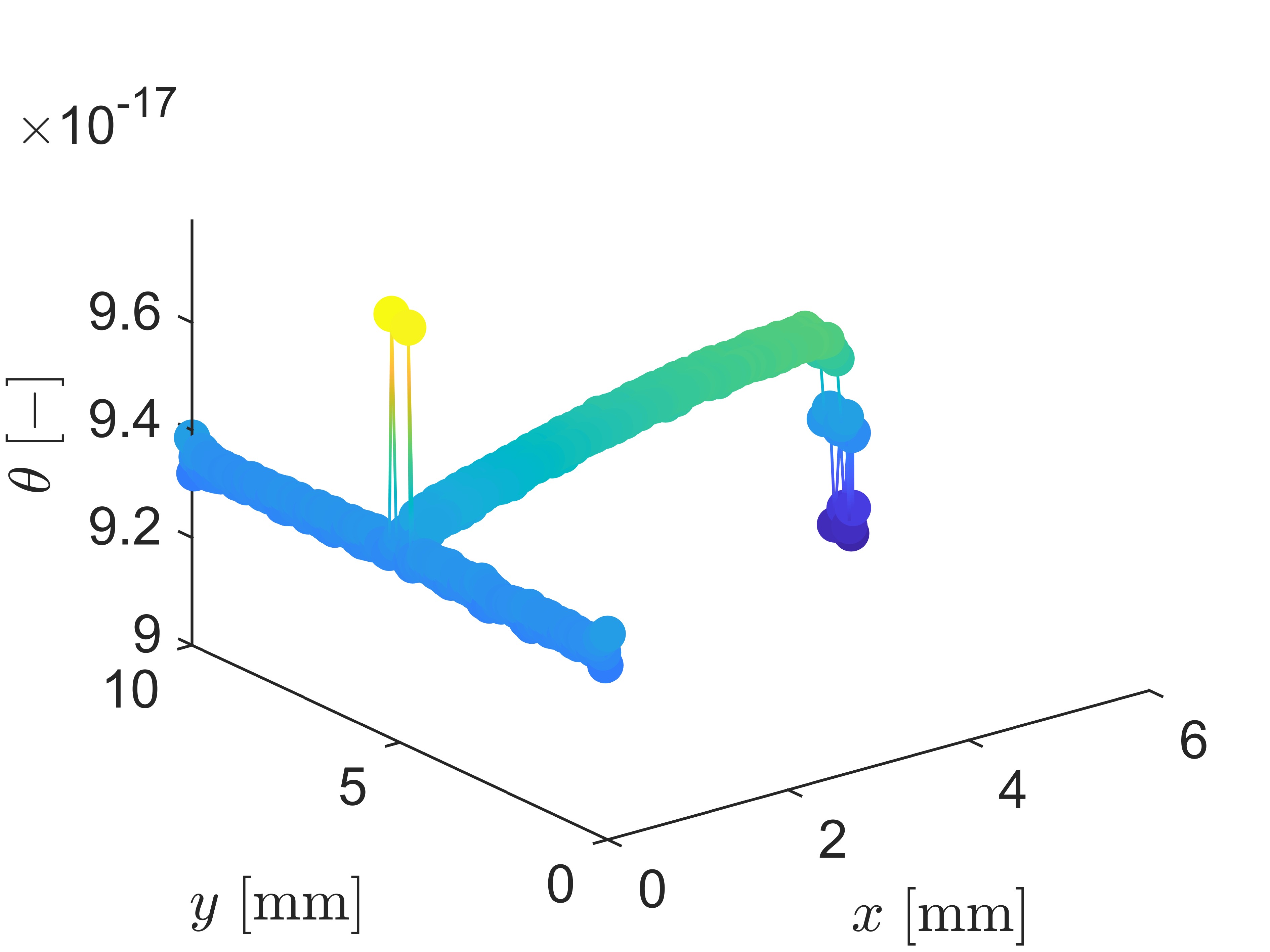}
         \caption{}
         \label{fig:effect_lumped_theta_PartiallyLumped}
     \end{subfigure}
     \begin{subfigure}{7.5cm}
        \vspace{1cm}
         \centering
         \textbf{Lumped, absorption and auto-ionisation}
         \includegraphics[width=7.5cm]{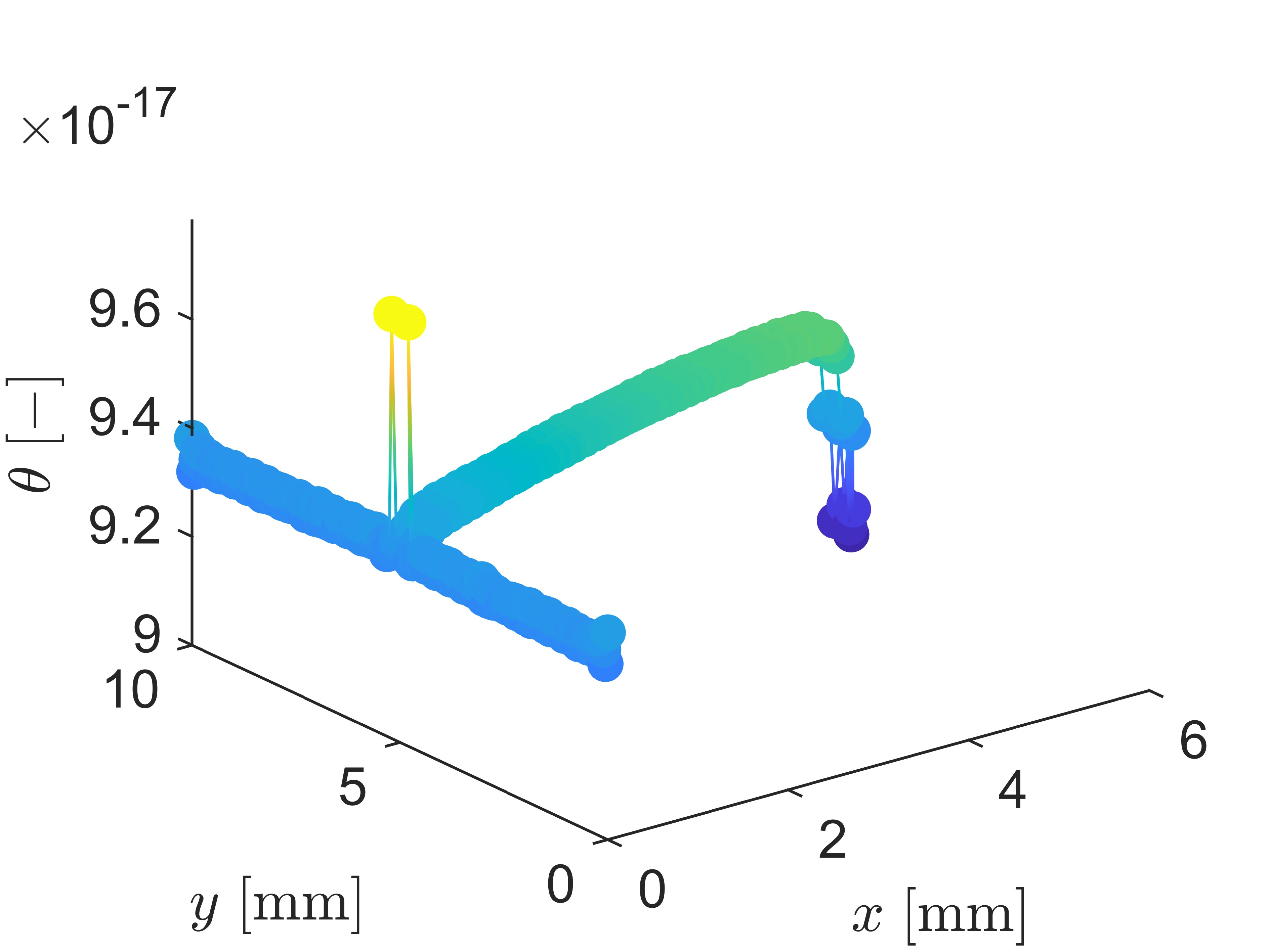}
         \caption{}
         \label{fig:effect_lumped_theta_FullLumped}
     \end{subfigure}
    \caption{Comparison of lumped and Gaussian integration by assessing their effect on the hydrogen surface coverage $\theta$ after a single iteration using $\Delta t = 1 \;\mathrm{s}$, and $k_4 = 10^3\;\mathrm{m}/\mathrm{s}$; (a) Gauss integration, (b) lumped integration for the absorption reaction, and (c) lumped integration for the absorption and auto-ionisation reactions.}
    \label{fig:effect_lumped_theta}
\end{figure}
\begin{figure}
    \centering
    \begin{subfigure}{7.5cm}
         \centering
         \textbf{Gauss}
         \includegraphics[width=7.5cm]{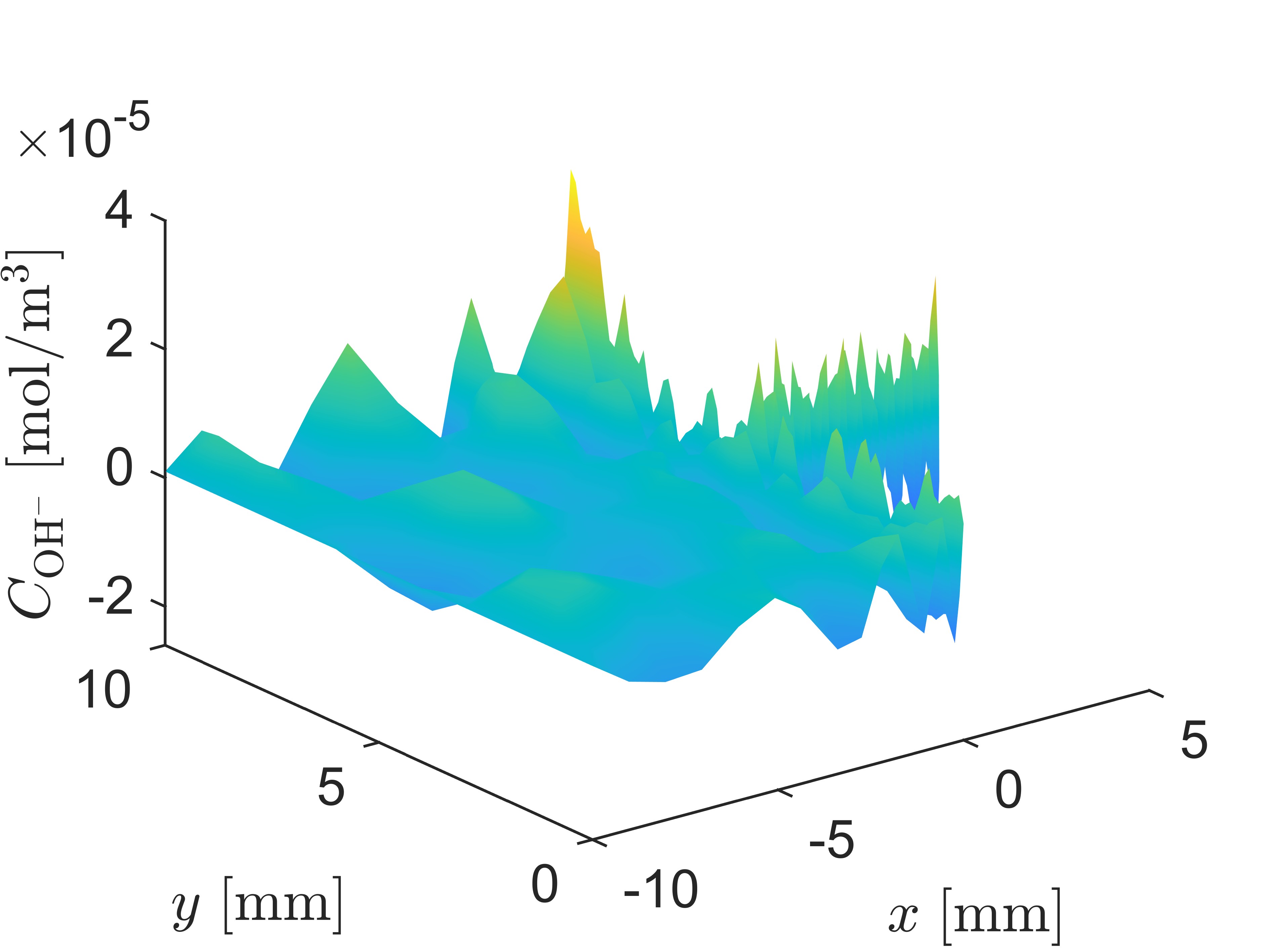}
         \caption{}
         \label{fig:effect_lumped_water_Standard}
    \end{subfigure}
    \begin{subfigure}{7.5cm}
         \centering
         \textbf{Lumped absorption}
         \includegraphics[width=7.5cm]{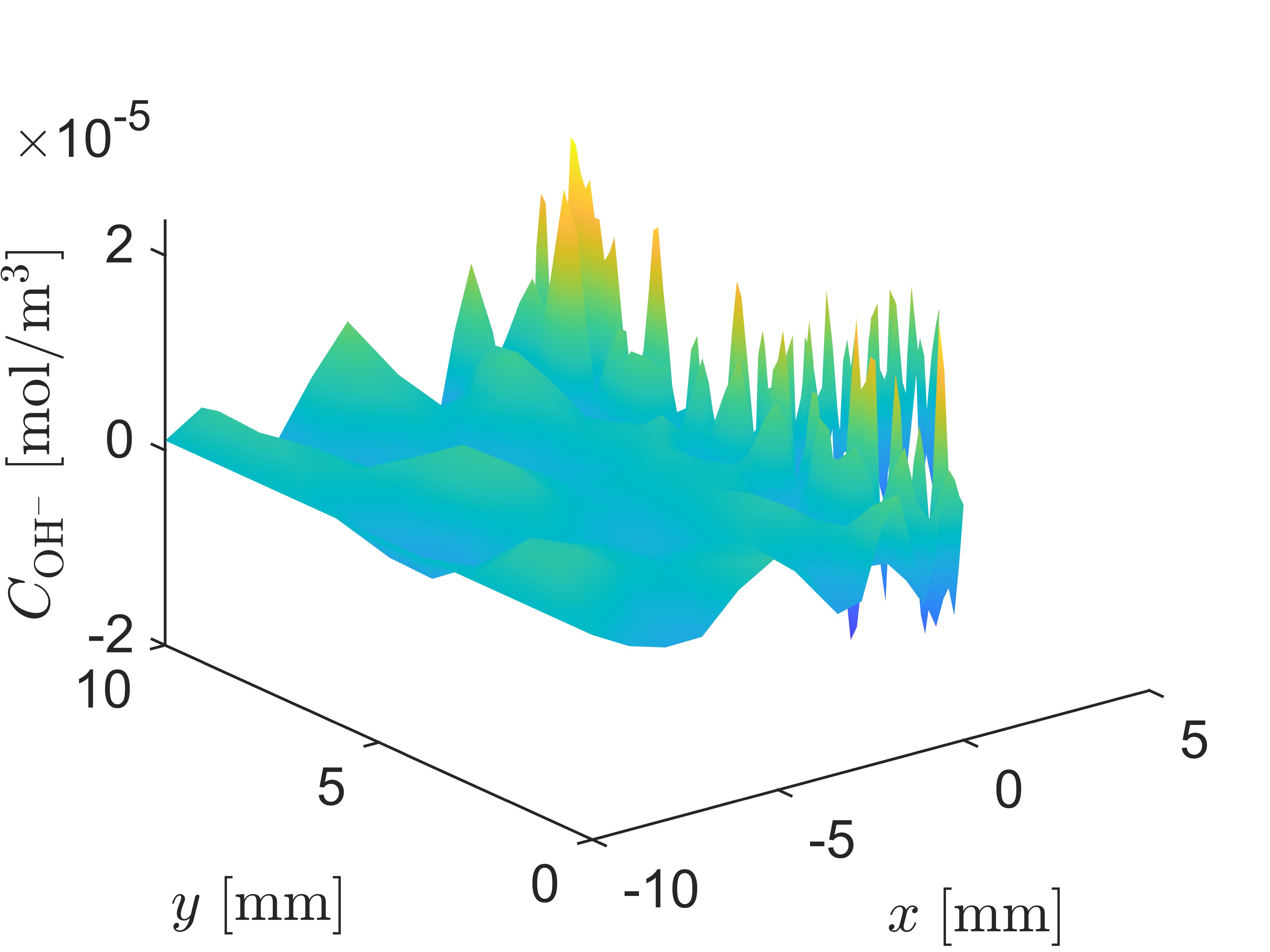}
         \caption{}
         \label{fig:effect_lumped_water_PartiallyLumped}
     \end{subfigure}
     \begin{subfigure}{7.5cm}
         \centering
         \vspace{1cm}
         \textbf{Lumped, absorption and auto-ionisation}
         \includegraphics[width=7.5cm]{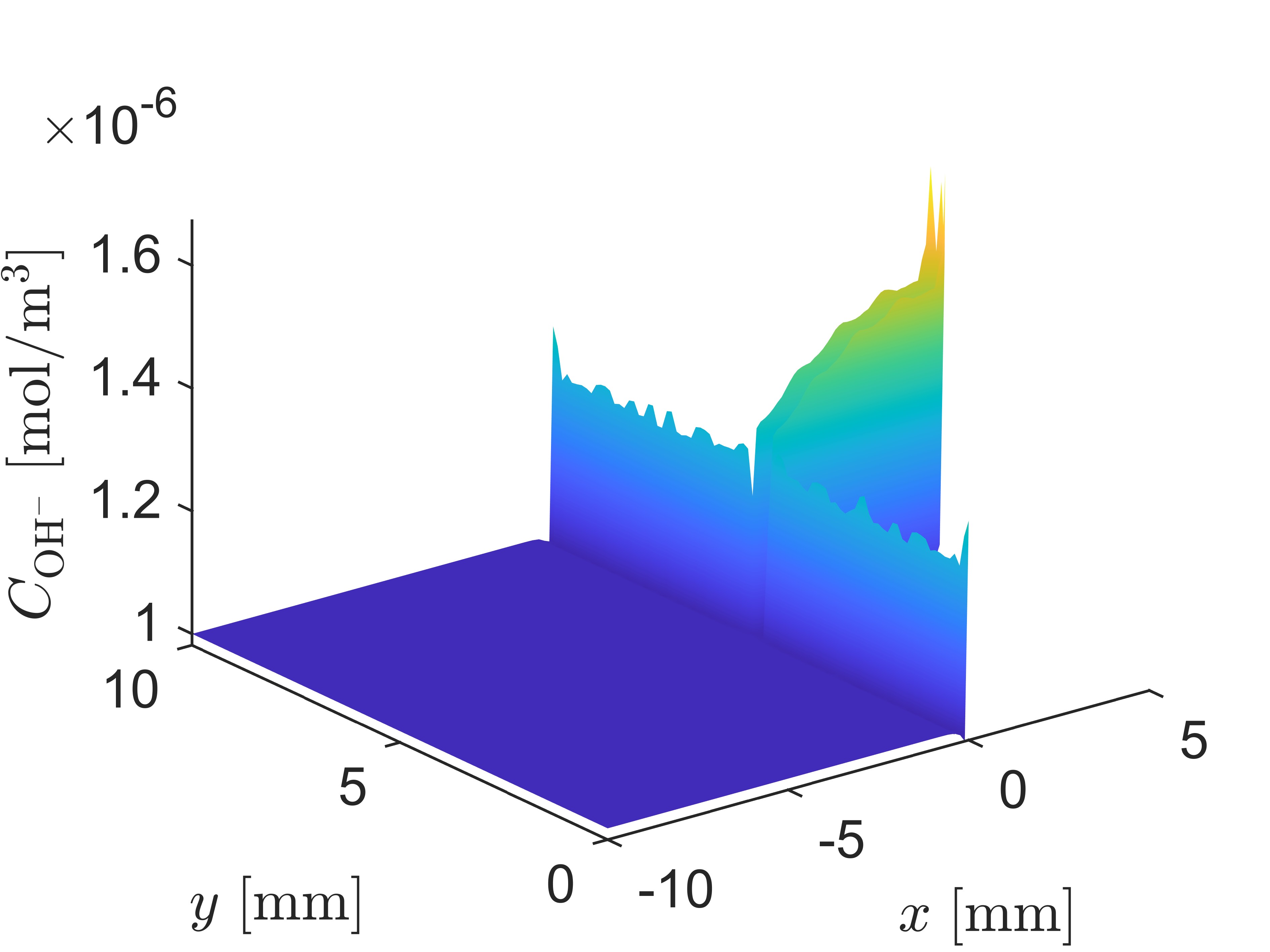}
         \caption{}
         \label{fig:effect_lumped_water_FullLumped}
     \end{subfigure}
    \caption{Comparison of lumped and Gaussian integration by assessing their effect on the $\mathrm{OH}^-$ concentration after a single iteration using $\Delta t = 1 \;\mathrm{s}$, and $k_4 = 10^3\;\mathrm{m}/\mathrm{s}$; (a) Gauss integration, (b) lumped integration for the absorption reaction, and (c) lumped integration for the absorption and auto-ionisation reactions.}
    \label{fig:effect_lumped_water}
\end{figure}

\begin{figure}
    \centering
    \includegraphics[width=12cm]{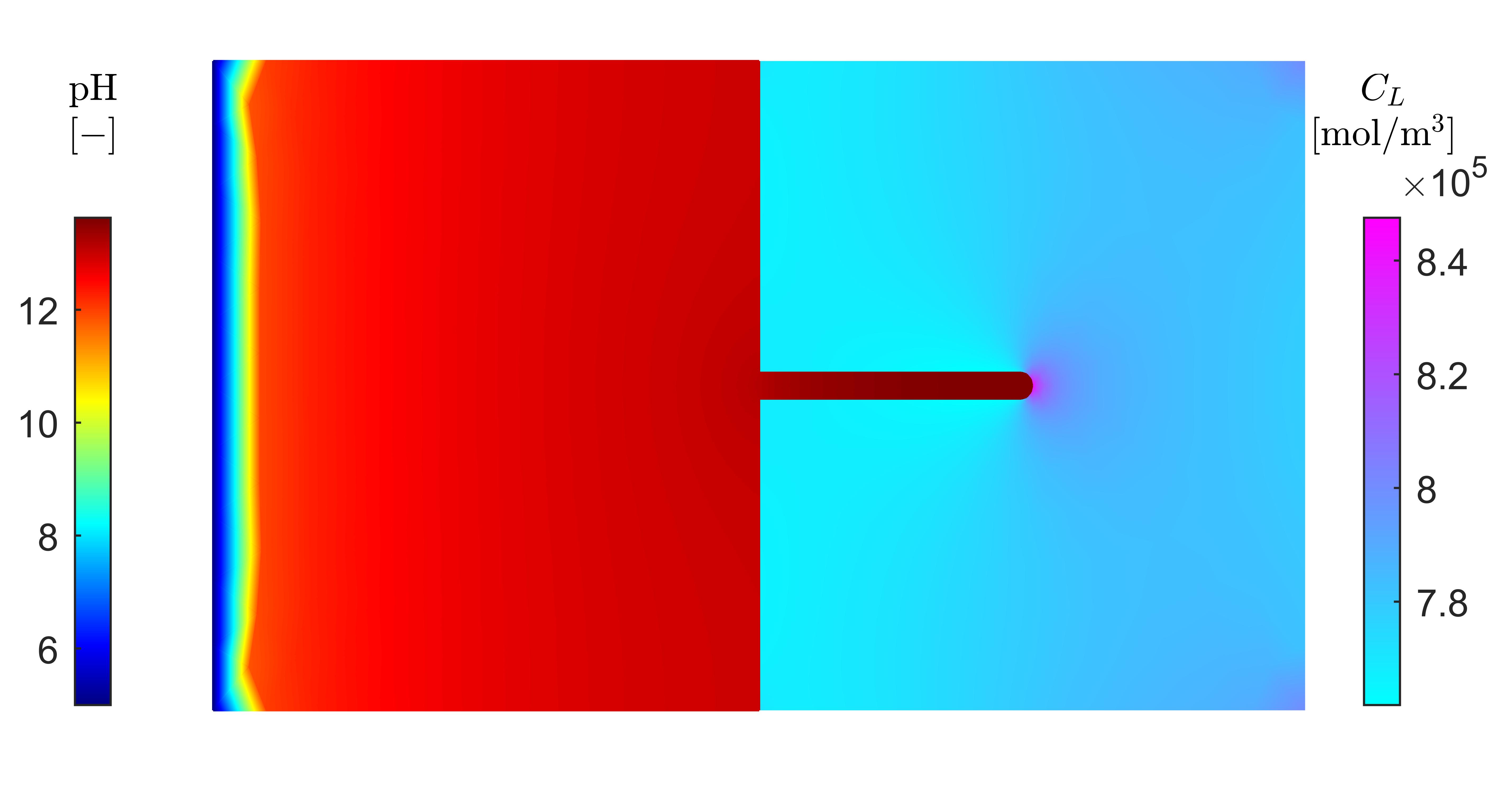}
    \caption{Contours of the electrolyte pH (left) and the lattice hydrogen concentration $C_L$ in the metal (right). The results are predicted for an applied potential of $E_m=-1\;\mathrm{V}_{\mathrm{SHE}}$ and at steady state ($t=50\;\mathrm{years}$).}
    \label{fig:CL_pH_Overview}
\end{figure}
\begin{figure}
    \centering
    \includegraphics[trim={0 0.5cm 0 0},clip]{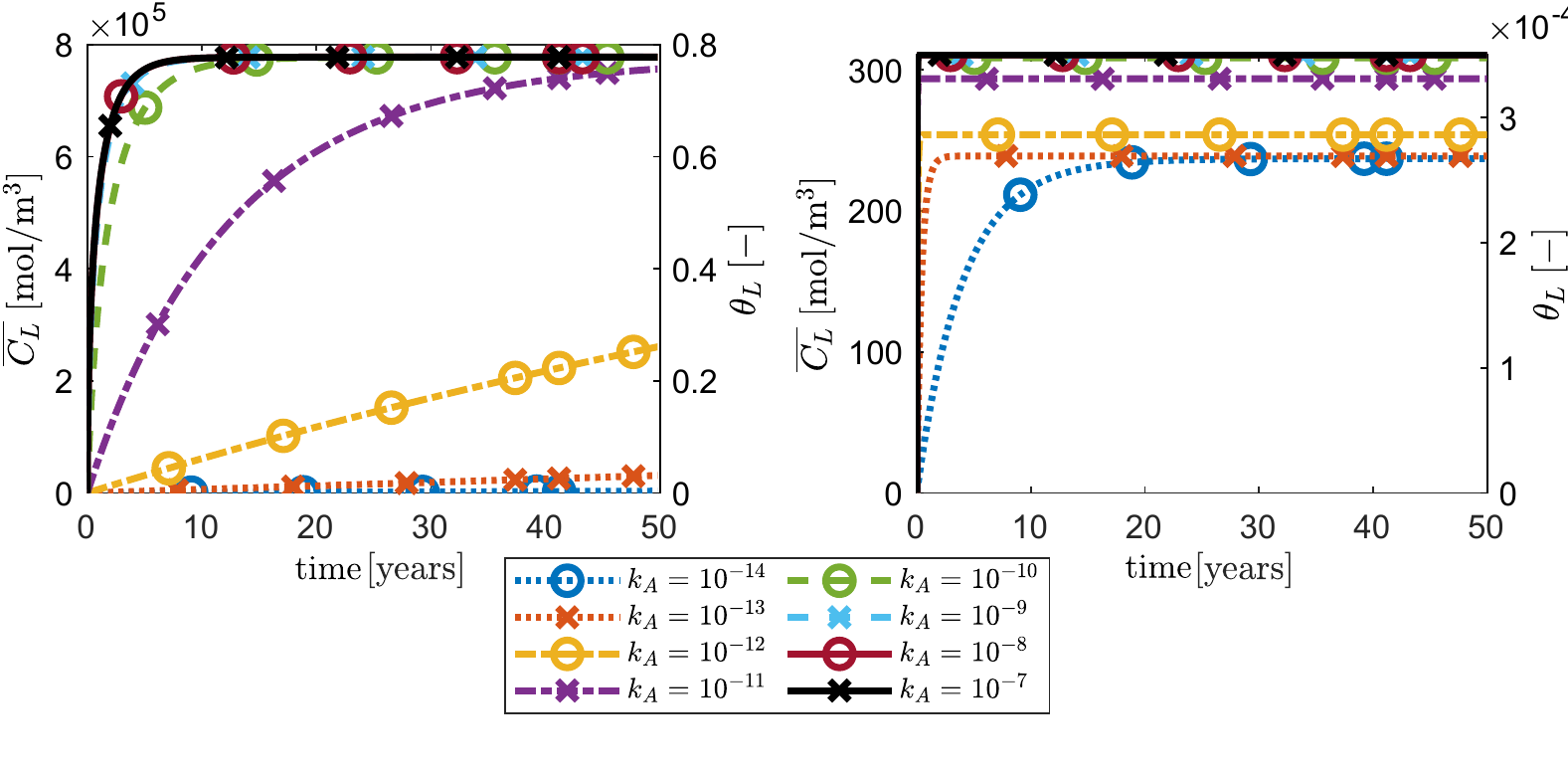}
    \begin{subfigure}{7.5cm}
         \centering
         \caption{$E_m=-1\;\mathrm{V}_{\mathrm{SHE}}$}
         \label{fig:effect_k4_vs_time_m1}
     \end{subfigure}
     \begin{subfigure}{7.5cm}
         \centering
         \caption{$E_m=0.5\;\mathrm{V}_{\mathrm{SHE}}$}
         \label{fig:effect_k4_vs_time_0}
     \end{subfigure}
    \caption{Influence of the absorption reaction rate constant $k_A\;[\mathrm{m}/\mathrm{s}]$ on the average lattice concentration over time.}
    \label{fig:effect_k4_vs_time}
\end{figure}
\begin{figure}
    \centering
    \begin{subfigure}{7.5cm}
         \centering
         \includegraphics[width=7.5cm]{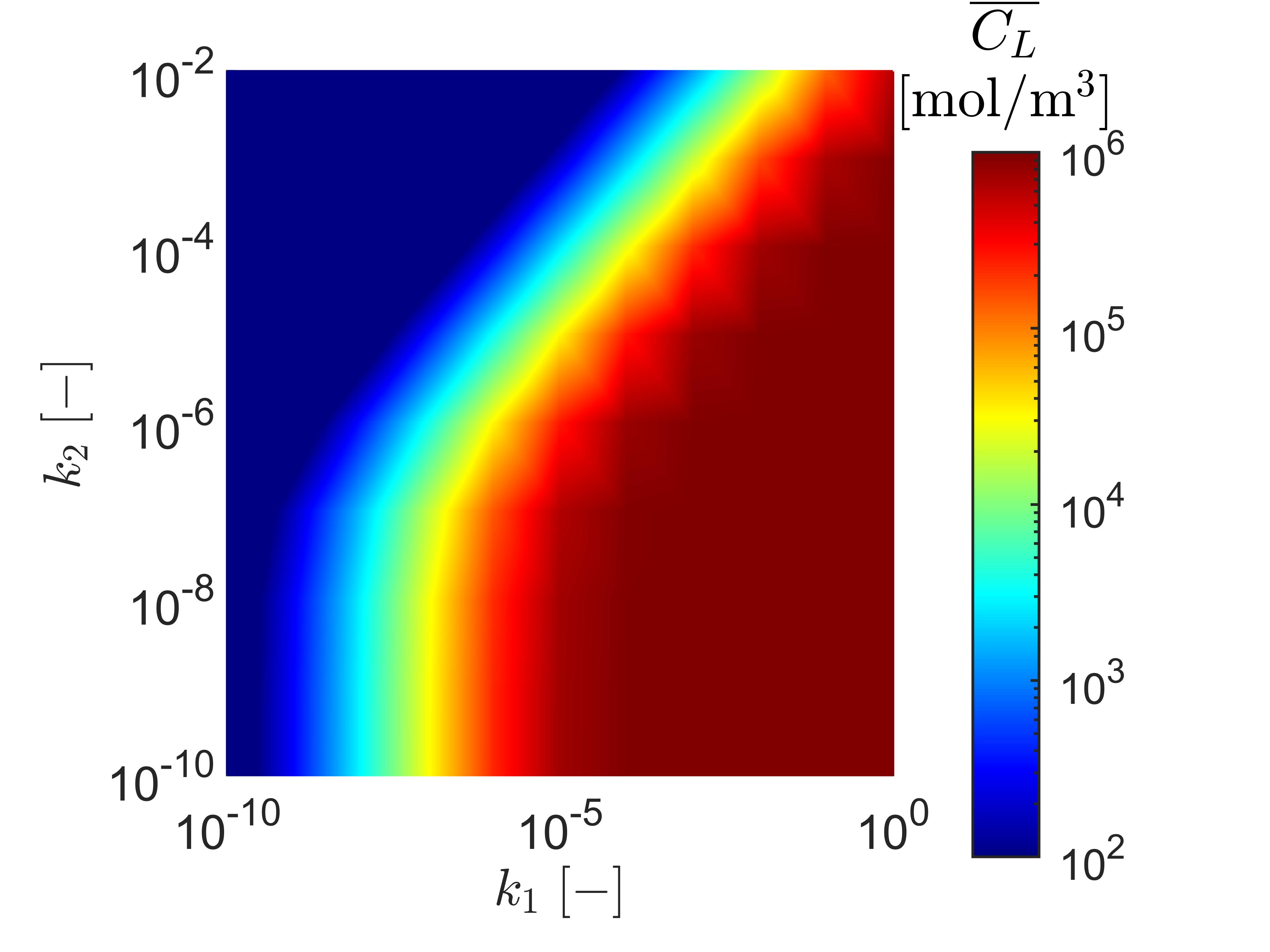}
         \caption{$E_m=-1\;\mathrm{V}_{\mathrm{SHE}}$}
         \label{fig:surface_k1_k2_em-1}
     \end{subfigure}
     \begin{subfigure}{7.5cm}
         \centering
         \includegraphics[width=7.5cm]{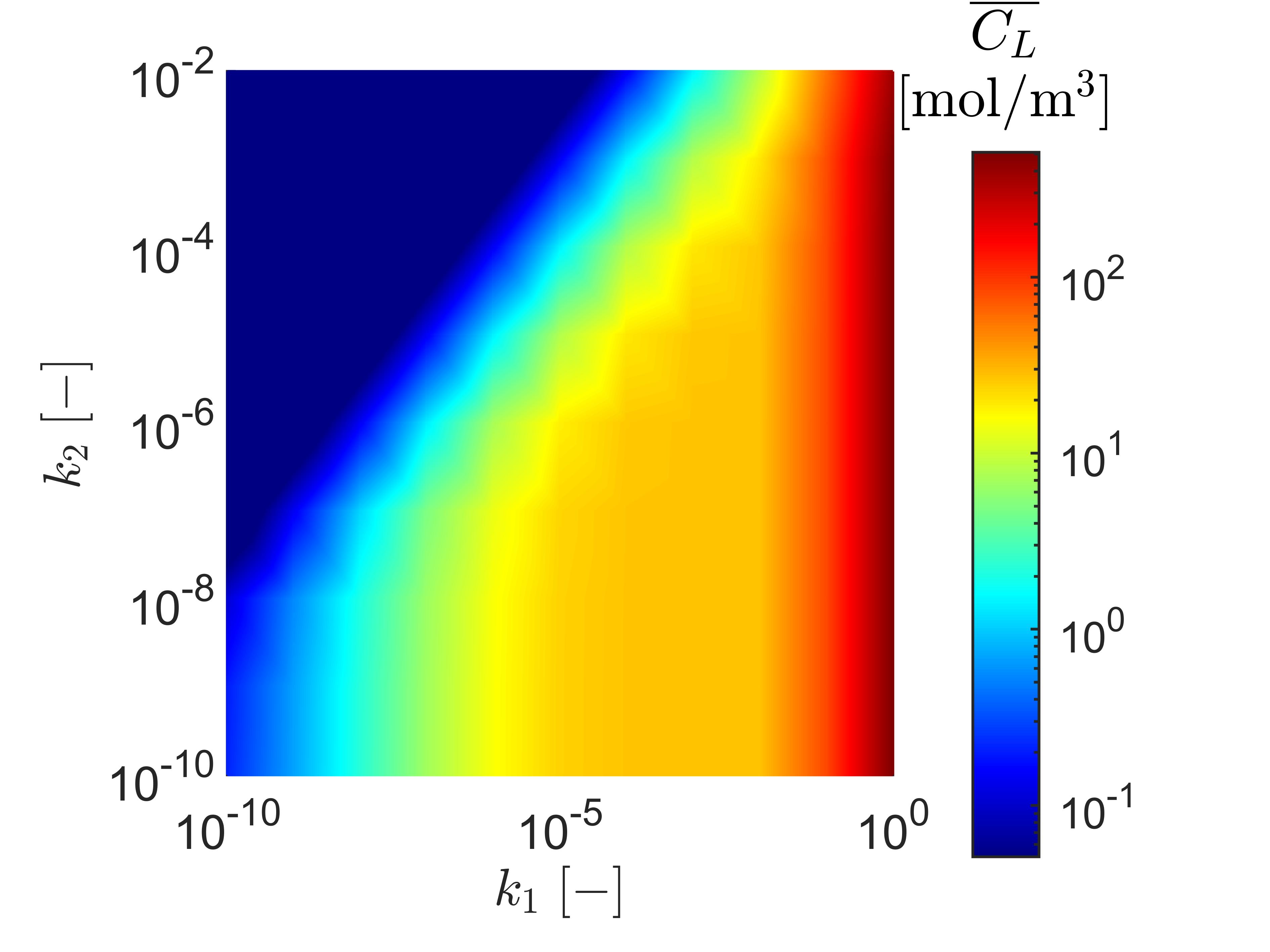}
         \caption{$E_m=0\;\mathrm{V}_{\mathrm{SHE}}$}
         \label{fig:surface_k1_k2_em0}
     \end{subfigure}
    \caption{Influence of the Volmer and Heyrovsky reaction rate constants on the average lattice hydrogen concentration $\overline{C}_L$ at $t=50\;\mathrm{years}$. As elaborated in the text, the reaction rate constant $k_1$ is defined to consistently vary the Volmer reaction rates, while the reaction rate constant $k_2$ refers to the Heyrovsky reaction rates.}
    \label{fig:surface_k1_k2}
\end{figure}
\begin{figure}
    \centering
        \begin{subfigure}{7.5cm}
         \centering
         \includegraphics[width=7.5cm]{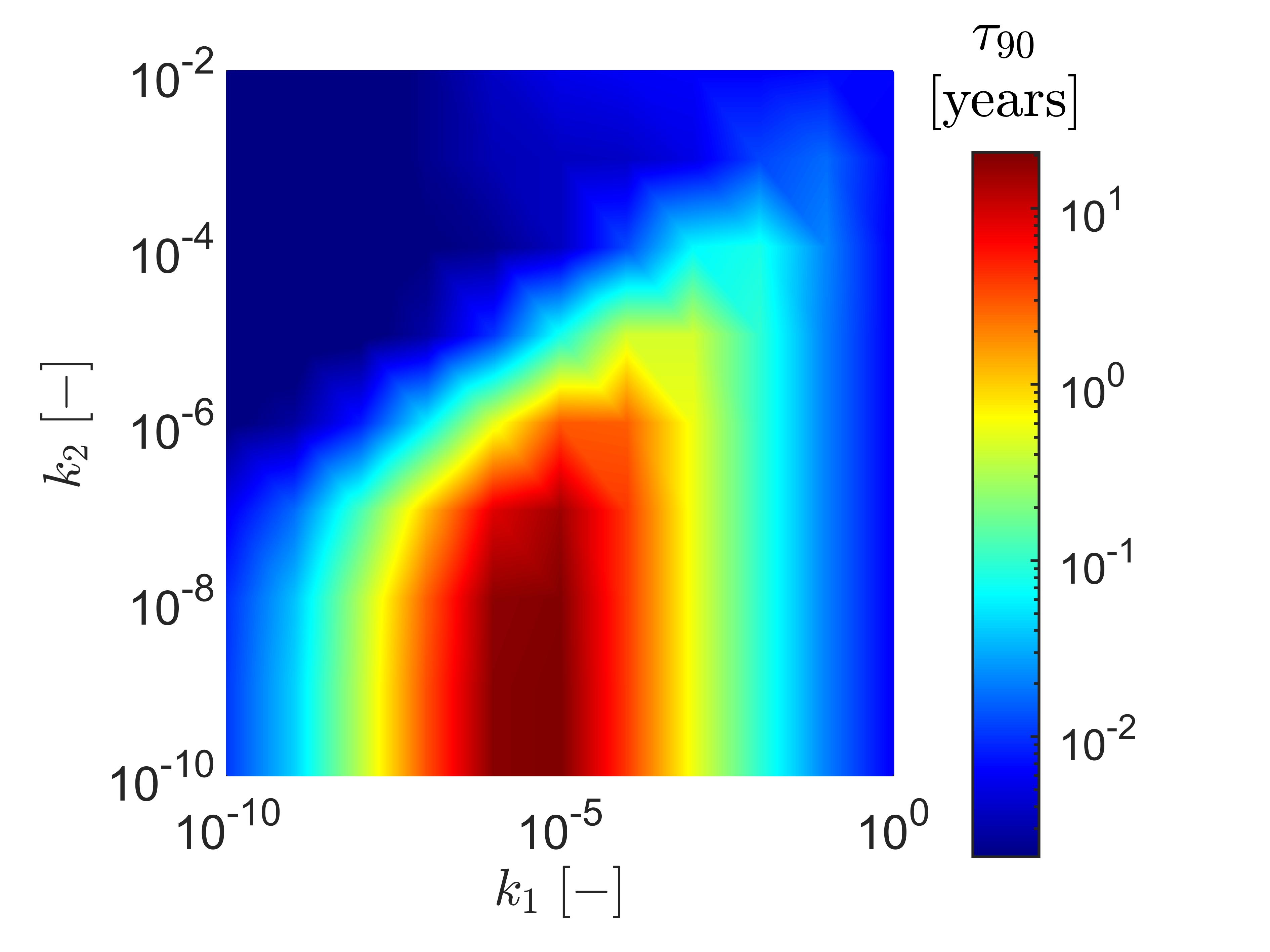}
         \caption{$E_m=-1\;\mathrm{V}_{\mathrm{SHE}}$}
         \label{fig:timesurface_k1_k2_em-1}
     \end{subfigure}
     \begin{subfigure}{7.5cm}
         \centering
         \includegraphics[width=7.5cm]{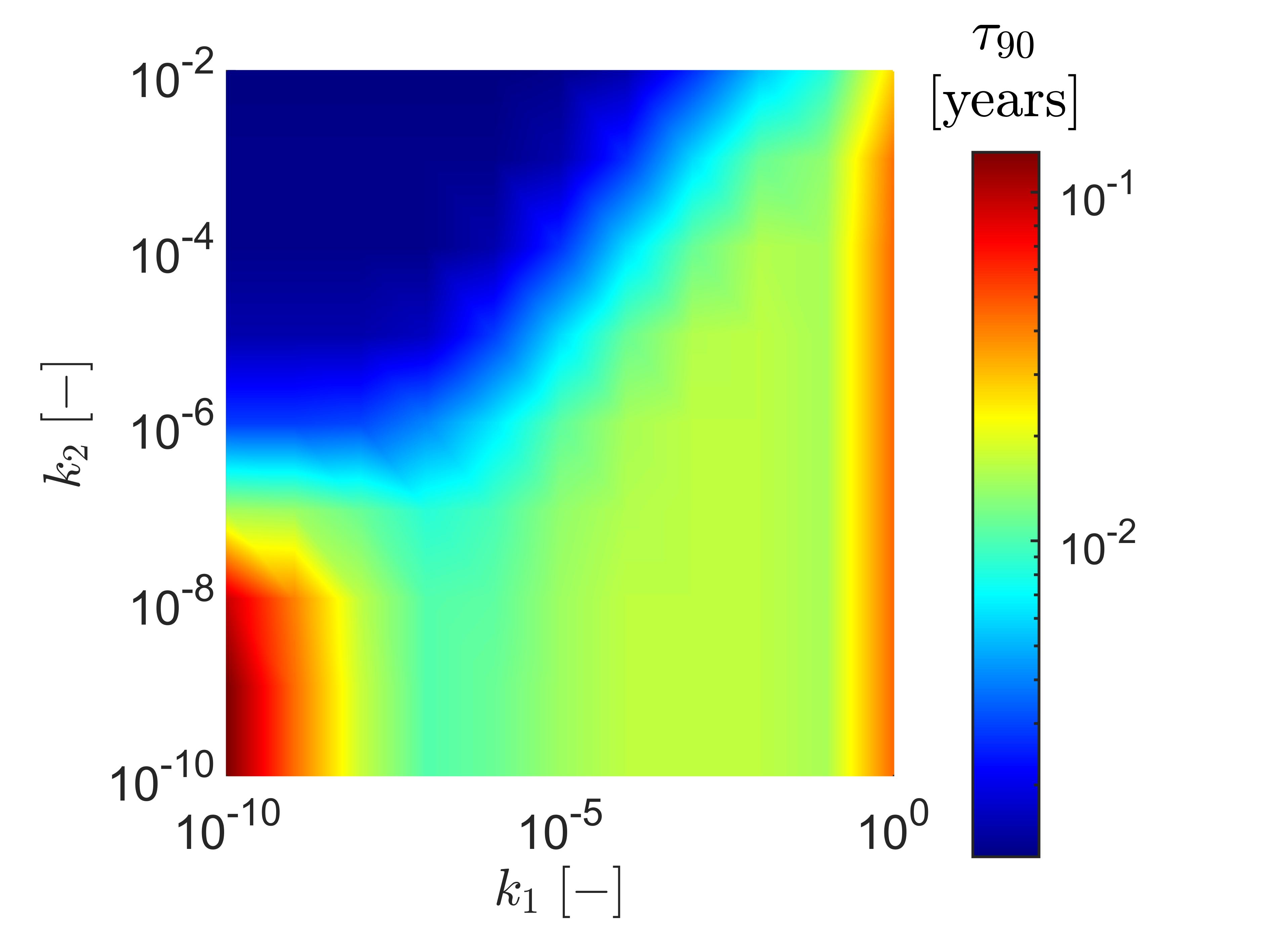}
         \caption{$E_m=0\;\mathrm{V}_{\mathrm{SHE}}$}
         \label{fig:timesurface_k1_k2_em0}
     \end{subfigure}
    \caption{Influence of the Volmer and Heyrovsky reaction rate constants on the time required to reach $90\%$ of the final (steady state) average lattice hydrogen concentration. As elaborated in the text, the reaction rate constant $k_1$ is defined to consistently vary the Volmer reaction rates, while the reaction rate constant $k_2$ refers to the Heyrovsky reaction rates.}
    \label{fig:timesurface_k1_k2}
\end{figure}

We proceed to demonstrate the potential and robustness of the lumped integration scheme presented above by simulating the uptake of hydrogen in the electrolyte-metal domain shown in \cref{fig:domain_overview}. First, the differences with Gauss integration are highlighted by examining the results obtained after a single iteration. Then, simulations are conducted for technologically-relevant ranges of material and environmental parameters, showcasing the ability of the lumped integration scheme-based framework to deliver predictions over previously unexplored conditions and time scales. In all cases, the electrolyte and metal sub-domains have $10\times10\;\mathrm{mm}$ dimensions, and a cracked region of $5\times0.4\;\mathrm{mm}$ is considered. Unless otherwise stated, the material and diffusion parameters are given in \cref{table:params}, while the reaction constants are listed in \cref{tab:reactionsused}. As initial condition, we use an electrolyte with $\mathrm{pH}=5$ and initial concentrations $C_{\mathrm{H}^+}=10^{-2}\;\mathrm{mol}/\mathrm{m}^3$, $C_{\mathrm{OH}^-}=10^{-6}\;\mathrm{mol}/\mathrm{m}^3$, $C_{\mathrm{Na}^+}=599.99\;\mathrm{mol}/\mathrm{m}^3$,  $C_{\mathrm{Cl}^-}=6\cdot10^{2}\;\mathrm{mol}/\mathrm{m}^3$, and $C_{\mathrm{Fe}^{2+}}=C_{\mathrm{FeOH}^+}=0\;\mathrm{mol}/\mathrm{m}^3$. Together with $\overline{\varphi}=0\;\mathrm{V}_{\mathrm{SHE}}$, these concentrations are also prescribed on the left edge as boundary conditions throughout the simulation. For the metal sub-domain, $C_L=0$ and $\mathbf{u}=\mathbf{0}$ are assumed at $t=0$ and a constant vertical displacement of $U_{ext} = 10\;\mathrm{\mu m}$ is prescribed at the top, while constraining both the horizontal and vertical displacements at the bottom. Finally, the only source of lattice hydrogen is through the metal-electrolyte interface, with all other boundaries not allowing any hydrogen flux. For the spatial discretisation, we use a finite element mesh composed of quadratic, triangular Bernstein elements with a characteristic element length of 0.1 mm near the metal-electrolyte interface and within the crack, increasing up to 0.5 mm further away. This results in a total of $3\cdot10^4$ degrees of freedom (DOFs).

\subsection*{Effectiveness of lumped integration} \label{sec:lumped_oscilllations_demo}

To showcase the effect of lumped integration, results are obtained after one single Newton-Raphson iteration and compared to those obtained using Gauss integration. More specifically, simulations are performed using no lumped integration, lumped integration for solely the absorption reaction, and lumped integration for both the absorption and the auto-ionisation reactions. The results obtained are shown in \cref{fig:effect_lumped_theta} for the surface coverage and in \cref{fig:effect_lumped_water} for the $\mathrm{OH}^-$ concentration. Consider first the results obtained for the surface coverage, \cref{fig:effect_lumped_theta}. Remarkable differences are observed between the results obtained with Gaussian and lumped integration, with the former showing severe oscillations and surface coverage values that are orders of magnitude off from the expected solution. While these are still non-converged results, and thus are not expected to be correct after a single iteration, the oscillations observed are indicative of convergence issues. However, when lumped integration is used, these oscillations disappear and the results obtained are in agreement with expectations after a single iteration: Attaining the right order of magnitude, solely positive surface coverage and concentrations, and tending towards the fully converged solution. Similar conclusions can be drawn from the $\mathrm{OH}^-$ concentration results, \cref{fig:effect_lumped_water}. Using Gauss integration and only using lumped integration for the absorption reaction results in oscillations an order of magnitude higher than the correct results, while using lumped integration for both reaction terms results in vastly improved results containing only minor oscillations. It should be noted that the removal of these oscillations does not guarantee an improved convergence radius and rate, but it is a good indication of such improvements. As such, these results emphasise the ability of the lumped integration scheme to enable the use of larger time increments and quantify hydrogen uptake over relevant parameter ranges without encountering convergence issues.

\subsection*{Predictions across environmental and material conditions}
\label{sec:sweepResults}

Converged results are subsequently obtained to demonstrate the ability of the lumped integration-based framework to map the parameter space and gain insight into hydrogen absorption behaviour over large time scales. All simulations are performed using an initial time increment of $\Delta t = 30\;\mathrm{s}$, increasing by $5\%$ per new time step until the full simulation duration of 50 years is reached after a total of 303 time steps. Within these $50\;\mathrm{years}$, all simulated cases achieve their steady-state solution, in which the hydrogen contents become constant. All results in this section presented as steady-state results are therefore the result of performing the time-dependent simulation for the full $50\;\mathrm{year}$ duration. By starting with a smaller time step, we accurately capture short-term behaviour (days) while exploiting the ever-increasing time-step size to efficiently obtain results for all relevant time scales in hydrogen embrittlement.More specifically, these time increments allow the rapid changes in $\mathrm{pH}$ and ionic concentrations to be captured, which occur on the order of minutes at the onset of the simulations. Increasing the increment afterwards allows modelling as well the slow hydrogen absorption timescales. A representative result from the converged simulations is shown in \cref{fig:CL_pH_Overview}, where contours of pH are shown on the electrolyte sub-domain while contours of lattice hydrogen concentration are presented on the metal sub-domain. The metal absorbs hydrogen from the neighbouring electrolyte, increasing the lattice hydrogen content in the material while decreasing the amount of hydrogen ions within the electrolyte and thus the pH. Within the metal, the largest concentration of hydrogen is located around the crack tip, having diffused to this location due to the large hydrostatic stresses present in this region. This boundary value problem was simulated in Ref. \cite{Hageman2022} using the commercial finite element package \texttt{COMSOL}. However, due to the described stability and oscillation issues inherent to Gaussian integration, \texttt{COMSOL}-based simulations were restricted to applied potentials equal to $E_m=-0.7\;\mathrm{V}_{\mathrm{SHE}}$ or higher, while here results are shown for $E_m=-1\;\mathrm{V}_{\mathrm{SHE}}$, a regime of relevance for hydrogen embrittlement and cathodic protection. Furthermore, the time step size restrictions required for stability when using Gauss integration (with time increments between $\Delta t=10^{-5}\;\mathrm{s}$ and $\Delta t=10^{-2}\;\mathrm{s}$) imposed a constraint on the time scales addressed. In particular, 82 hours were needed to simulate 10 min of hydrogen uptake in an Intel i7-10700 CPU. In contrast, the described lumped integration scheme, not suffering from these limitations, is able to deliver predictions over a time scale of 50 years within 3 hours, using the same system.\\

Taking advantage of the robustness and stability of the lumped integration implementation, we proceed to obtain results in regimes previously unexplored yet critical for hydrogen embrittlement predictions. To this end, the absorbed hydrogen is quantified using the average and maximum lattice hydrogen concentrations, $\overline{C}_L$ and $C_L^{max}$ respectively. The maximum concentration is always located at the crack tip, due to stress localisation, whereas the average is obtained by numerically integrating the lattice hydrogen over the complete metal subdomain, and normalising it with the metal surface area. These two quantities give a good indication of the total behaviour of the system; the average hydrogen concentration provides an estimate of overall hydrogen uptake, and the maximum hydrogen concentration indicates how unevenly hydrogen is distributed due to the presence of the stresses within the metal. 

\subsubsection*{Application across the reaction constant space}
\label{sec:constants_sweep}

We begin to explore the reaction constant space by varying the values of $k_{A}$ and $k_{A}'$ within the experimentally reported range. All other reaction rate constants are kept constant and take the values given in Table \ref{tab:reactionsused}. The experimental literature reports values for $k_{A}$ spanning the range $k_{A}=2.4\cdot10^{-12}\;\mathrm{m}/\mathrm{s}$ \cite{Elhamid2000b} to $k_{A}=1.2\cdot10^5\;\mathrm{m}/\mathrm{s}$ \cite{Turnbull1996}. Accordingly, simulations are conducted within the range $k_{A}=[10^{-14}, 10^5]\;\mathrm{m}/\mathrm{s}$. The backward reaction constant is chosen such that $k_A'/k_A=7\cdot10^4$, altering the rate at which the two reactions occur but not their equilibrium. These simulations are performed at four applied potentials, $E_m = -1\;\mathrm{V}_{\mathrm{SHE}}$, $E_m = -0.5\;\mathrm{V}_{\mathrm{SHE}}$, $E_m = 0\;\mathrm{V}_{\mathrm{SHE}}$, and $E_m = 0.5\;\mathrm{V}_{\mathrm{SHE}}$, going from a hydrogen-dominated regime to a corrosion reaction dominated one. The resulting hydrogen uptake within the metal is shown in \cref{fig:effect_k4_vs_time} using $E_m=-1\;\mathrm{V}_{\mathrm{SHE}}$. The results are only shown up to $k_A=10^{-7}\;\mathrm{m}/\mathrm{s}$, as higher absorption reaction constants led to almost identical predictions. Changes in the absorption reaction constant are seen to strongly affect the rate at which the lattice hydrogen concentration achieves equilibrium with the surrounding electrolyte. For unrealistically low values of $k_A$ ($\sim 10^{-14}$), very little hydrogen enters the metal within the simulated 50 years. In contrast, values of $k_A=10^{-9}\;\mathrm{m}/\mathrm{s}$ and higher lead to the same average lattice hydrogen concentrations, indicating that the hydrogen absorption is almost instantaneous, and limited either by the diffusion within the metal or by the other surface reactions. Results obtained using positive metal potentials are shown in \cref{fig:effect_k4_vs_time_0} using $E_m=0.5\;\mathrm{V}_{\mathrm{SHE}}$. Hydrogen uptake is hindered by the inhibiting effect of the metal potential, leading to a lower level of absorbed hydrogen and thus a shorter time needed to achieve steady state. No sensitivity to the absorption rate constant is observed for high $k_A$ values. However, as $k_A$ takes lower values, small differences are observed in the magnitude of the lattice hydrogen concentration at steady state. This points to a relation between the equilibrium hydrogen surface coverage and the absorption rate constant for low $k_A$ values, as the Heyrovsky and Tafel reactions become dominant and remove hydrogen faster than it is absorbed, resulting in a lower degree of hydrogen uptake.\\ 

The role of the Volmer and Heyrovsky reaction constants is subsequently investigated. Two reaction rate constants ($k_1$, $k_2$) are introduced to facilitate interpretation of the results and ensure consistency across the acid and basic regimes - reaction rate constants are chosen to ensure that acidic reactions are dominant below $\mathrm{pH}=7$ while basic reactions dominate above $\mathrm{pH=7}$. The reaction rate constant $k_1$ alters the Volmer reaction rates, while $k_2$ varies the Heyrovsky rate. From these two constants, we construct the reaction constants used in our model as: $k_{Va}=k_1$, $k_{Va}'=k_1\cdot10^{-6}$, $k_{Vb}=k_1\cdot 10^{-4}$, $k_{Vb}'=k_1\cdot10^{-9}$, $k_{Ha}=k_2$, and $k_{Hb} = k_2\cdot10^{-4}$. The values for $k_1$ are varied between $10^{-10}$ and $1$, while $k_2$ is varied between $10^{-10}$ and $10^{-2}$, covering the complete range of values reported in literature \citep{Hageman2022}. The average lattice hydrogen concentrations obtained in the metal while varying these two parameters are shown in \cref{fig:surface_k1_k2}. Two applied potentials are considered, $E_m = -1\;\mathrm{V}_{\mathrm{SHE}}$ and $E_m = 0\;\mathrm{V}_{\mathrm{SHE}}$. For both cases, very low $k_1$ values result in negligible hydrogen uptake, independently of the value of $k_2$. For increased rates for the Volmer reaction (higher $k_1$), more hydrogen is present within the metal. This effect is strongest for the negative metal potential, where the Volmer reaction is further accelerated due to the role of the electric overpotential. In contrast, increasing the Heyrovsky reaction rates (higher $k_2$) reduces the amount of hydrogen available. As a result, the metal is nearly saturated when high Volmer reaction constants are combined with low Heyrovsky reaction constants. The time required to attain a hydrogen content that is 90\% of the steady state value is shown in \cref{fig:timesurface_k1_k2} as a function of $k_1$ and $k_2$. For the $E_m=-1\;\mathrm{V}_{\mathrm{SHE}}$ simulations, the longest required times correspond to the highest hydrogen contents, requiring close to $15\;\mathrm{years}$ to attain these steady state values for $k_1 \approx 10^{-5}$ and low $k_2$ values. When the Volmer reaction rates are decreased by lowering $k_1$, the time required to reach steady state is strongly reduced to only several days. In contrast, increasing the Volmer rates beyond $k_1=10^{-5}$ brings in a reduction in the time needed to achieve steady state. This corresponds with the reaction rates at which the metal becomes fully saturated, and thus all additional hydrogen produced at the surface is no longer used to increase the lattice concentration, but rather spent on reaching the equilibrium state. For the $E_m=0\;\mathrm{V}_{\mathrm{SHE}}$ simulations, less hydrogen is present in the metal lattice at equilibrium, and as such less time is required to attain this equilibrium.\\

From the results presented in this section, it can be concluded that the presented scheme is stable, oscillation-free, and well-converging across the whole range of relevant reaction constants. Furthermore, it is interesting to note that the range of values reported in the literature causes a significant spread in the results, ranging from close to no lattice hydrogen to a fully saturated metal.

\subsubsection*{Application across the metal potential space}
\label{sec:metal_potential}
\begin{figure}
    \centering
    \includegraphics[trim={0 1cm 0 0},clip]{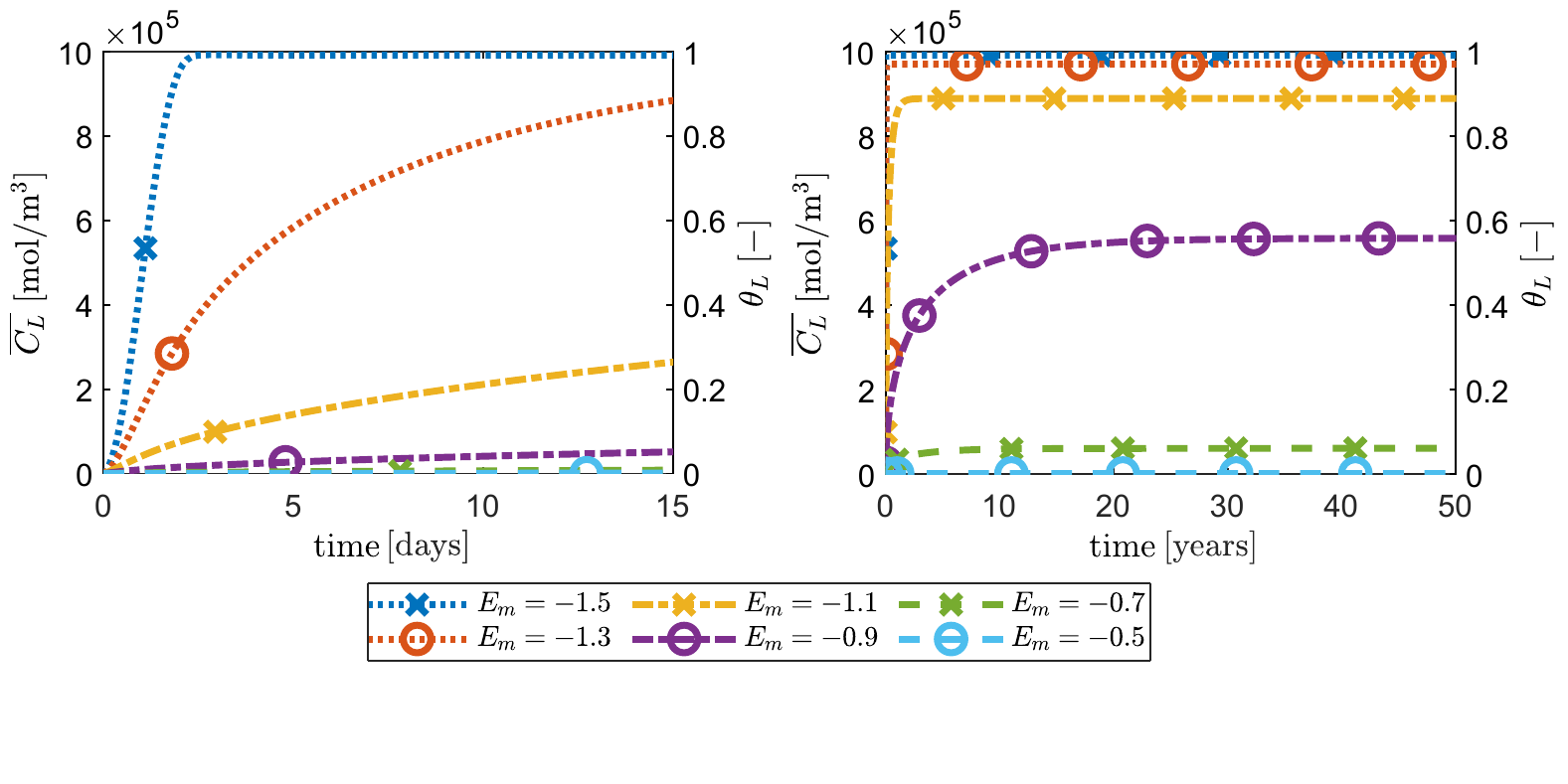}
    \begin{subfigure}{7.5cm}
         \centering
         \caption{}
         \label{fig:Em_vs_CLaverage_nonZoom}
    \end{subfigure}
    \begin{subfigure}{7.5cm}
         \centering
         \caption{}
         \label{fig:Em_vs_CLaverage_zoom}
     \end{subfigure}
    \caption{Influence of the applied potential $E_m$ (in $\mathrm{V}_{\mathrm{SHE}}$) on hydrogen uptake; evolution of the average lattice hydrogen concentration $\overline{C}_L$ (a) over the first 15 days, and (b) over 50 years.}
    \label{fig:Em_vs_CLaverage}
\end{figure}
\begin{figure}
    \centering
    \includegraphics{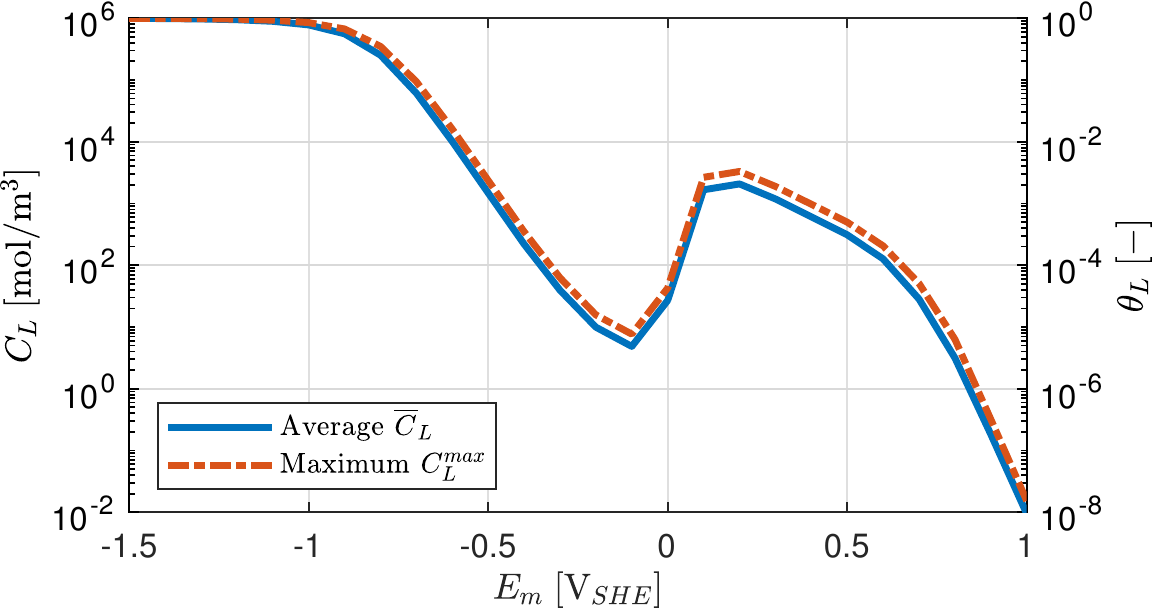}
    \caption{Influence of the applied potential $E_m$ on hydrogen uptake; steady state predictions of lattice hydrogen concentration (left) and occupancy (right). Results are shown for the maximum ($C_L^{max}$, orange dashed line) and average ($\overline{C}_L$, blue solid line) hydrogen concentrations at equilibrium.}
    \label{fig:Em_vs_CLmax}
\end{figure}

\begin{figure}
    \centering
    \includegraphics{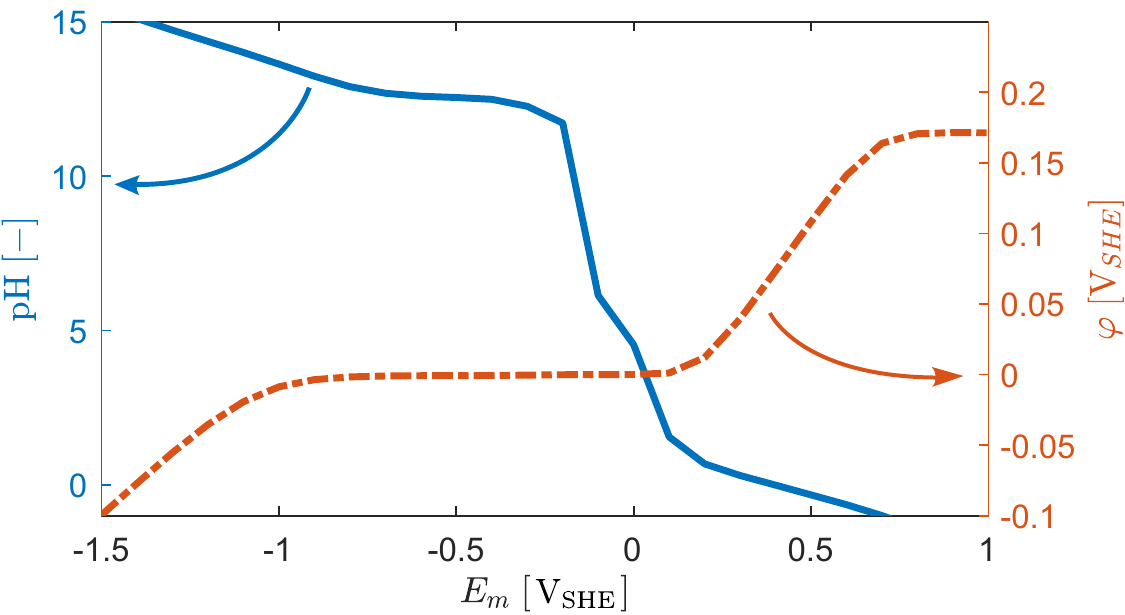}
    \caption{Sensitivity of the environmental conditions at the crack tip to the applied potential $E_m$. The left axis shows the change in electrolyte pH (blue solid curve), while the right axis denotes the variation in electrolyte potential $\varphi$ (red dashed curve).}
    \label{fig:pHphi_vs_Em}
\end{figure}

Next, we investigate the influence of the applied potential on hydrogen uptake, keeping the reaction rate parameters constant. The enhanced stability provided by the lumped integration scheme presented here enables obtaining results over the range $E_m =[ -1.5, 1] \;\mathrm{V}_{\mathrm{SHE}}$, going from strong cathodic protection conditions to the corrosion dominated regime. Previous reported results were limited to the range $E_m =[ -0.7, 0.5] \;\mathrm{V}_{\mathrm{SHE}}$, due to convergence issues resulting from the above discussed stability problems \cite{Hageman2022}. The evolution in time of the average content of hydrogen within the metal is shown in \cref{fig:Em_vs_CLaverage}. When the applied potential is $-1.5\;\mathrm{V}_{\mathrm{SHE}}$, the metal becomes fully saturated within two days. Strong negative potentials accelerate surface reaction, leading to rapid saturation of the surface sites and the metal bulk sites near the surface. This, in turn, enhances hydrogen diffusivity within the metal through the non-linear diffusion term, see \cref{eq:massbalance_lattice}. This high diffusivity facilitates the distribution of hydrogen within the specimen, more readily attaining steady state conditions. On the other hand, increasing the metal potential brings in a reduction in electric overpotential, and thus decreases the forward surface reaction rates while accelerating the backward reaction rates. As a result, for the $E_m=-1.3\;\mathrm{V}_{\mathrm{SHE}}$ simulations it takes over $20\;\mathrm{days}$ to attain steady state, with the lattice concentration at this steady state being slightly lower compared to the $E_m=-1.5\;\mathrm{V}_{\mathrm{SHE}}$ results. Further increasing the metal potential shows a more pronounced effect, with hydrogen-producing surface reactions becoming slower and the steady state $\overline{C}_L$ becoming smaller, which results in a shorter time needed to attain this equilibrium condition.\\ 

The relationship between the lattice hydrogen concentration and the applied potential at steady state is shown in \cref{fig:Em_vs_CLmax}, using a logarithmic scale. Both the average ($\overline{C}_L$) and maximum ($C_L^{max}$) lattice hydrogen concentrations are provided. This mapping is expected to be useful for the hydrogen assisted fracture community, as it provides a first order approximation for input of chemo-mechanical models of hydrogen embrittlement aiming at delivering predictions over large time scales. The results show that at strongly negative potentials the average and maximum concentrations are closer to each other since the role of stress raisers is reduced, as discussed in the context of \cref{eq:CMAX}. Increasing the metal potential lowers both the average and maximum concentrations. Going into positive metal potentials initially increases the hydrogen uptake through an increase in corrosion rate and subsequent reduction in pH. Increasing the metal potential even further to $E_m=1\;\mathrm{V}_{\mathrm{SHE}}$ prevents any of the hydrogen reactions from occurring due to the high electric overpotential, even though the pH of the electrolyte is strongly acidic. This is also seen in \cref{fig:pHphi_vs_Em}, showing the pH and electrolyte potential within the crack. At low metal potentials, the environment becomes highly basic and the electrolyte potential decreases. The opposite happens for high potentials, increasing the metal potential while lowering the pH. Here, one should note that due to the limited size of the domain and the zero electric potential boundary condition on the left boundary, the electrolyte never attains the same potential as the metal. This explains the ever-accelerating effect of decreasing the potential on the hydrogen reactions and the strongly inhibiting effect at high potentials.

%\FloatBarrier
\section*{Application to tensile rods contained within an electrolyte}
\label{sec:3D_cases}
\begin{figure}
    \centering
    \includegraphics{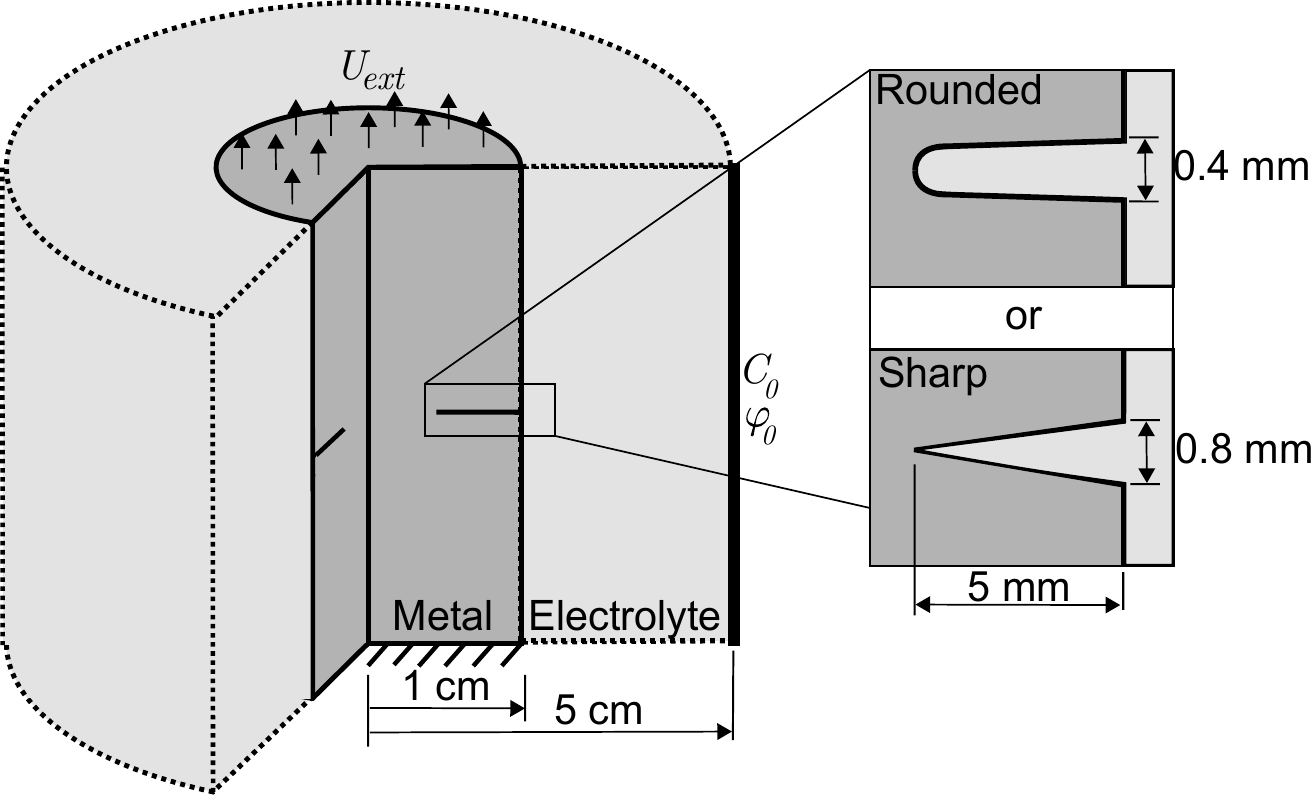}
    \caption{Domains considered for the tensile rod cases, containing either a rounded crack or a sharp crack.}
    \label{fig:domain_3D}
\end{figure}
\begin{figure}
    \centering
    \begin{subfigure}{7.5cm}
         \centering
         \includegraphics[width=7.5cm]{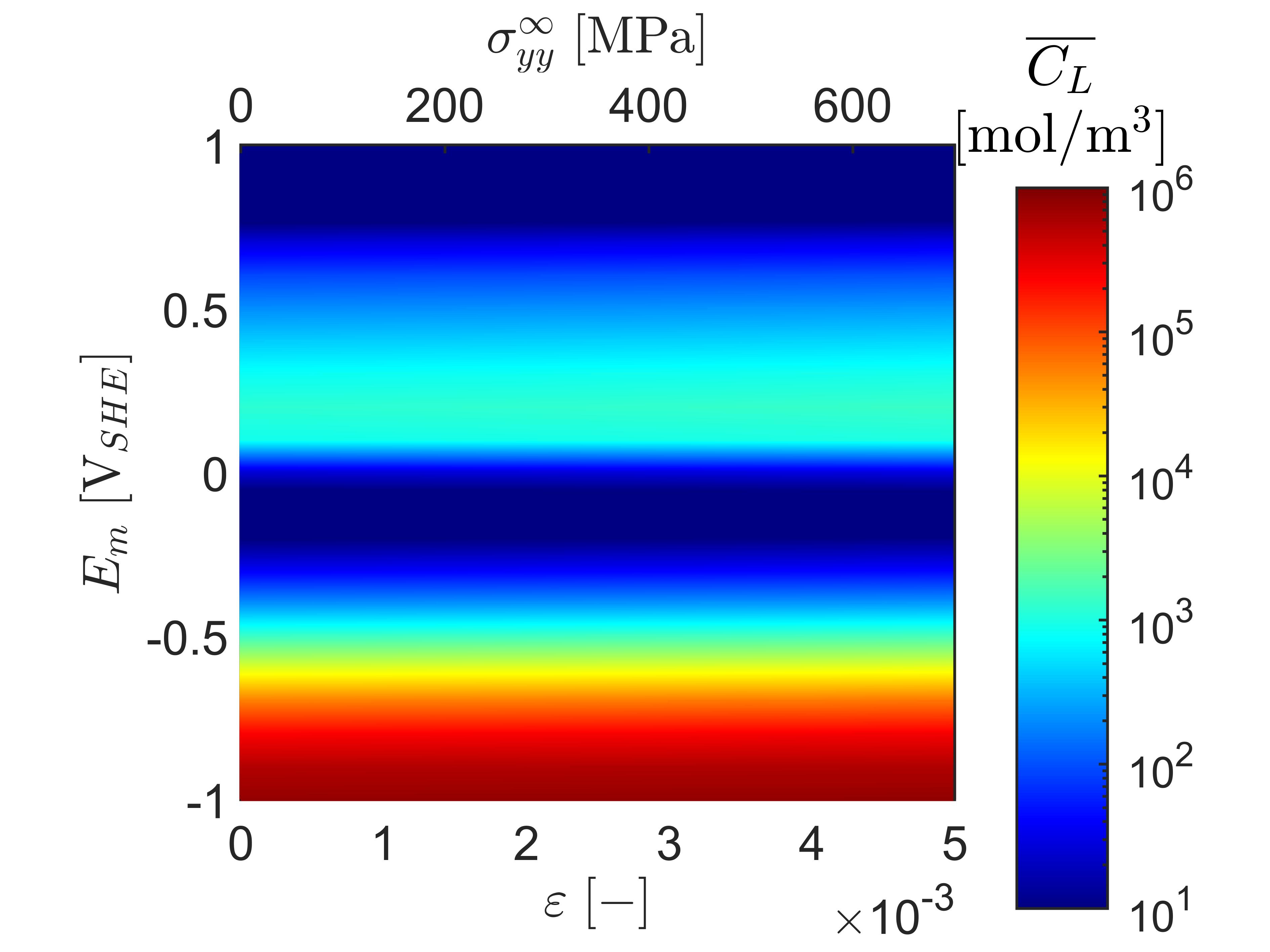}
         \caption{}
         \label{fig:rounded_CL_Average}
    \end{subfigure}
    \begin{subfigure}{7.5cm}
         \centering
         \includegraphics[width=7.5cm]{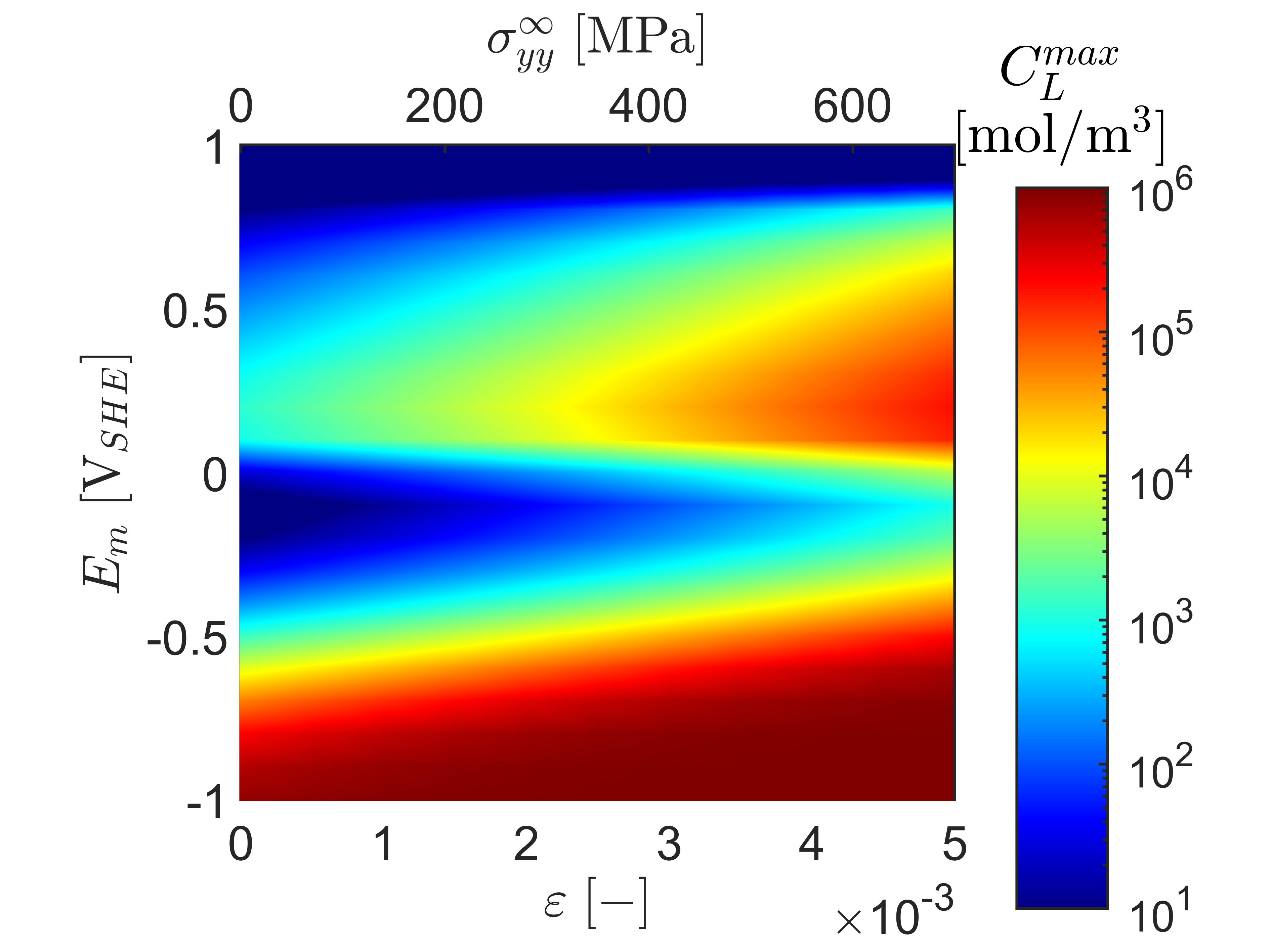}
         \caption{}
         \label{fig:rounded_CL_Max}
     \end{subfigure}
    \caption{Hydrogen uptake sensitivity to mechanical straining. Steady state results for the blunted crack case showing: (a) the average lattice hydrogen concentration $\overline{C}_L$, and (b) and the maximum hydrogen concentration $C_L^{max}$ as a function of the remote strain/stress and the applied potential $E_m$.}
    \label{fig:rounded_CL}
\end{figure}
\begin{figure}
    \centering
    \begin{subfigure}{7.5cm}
         \centering
         \includegraphics[width=7.5cm]{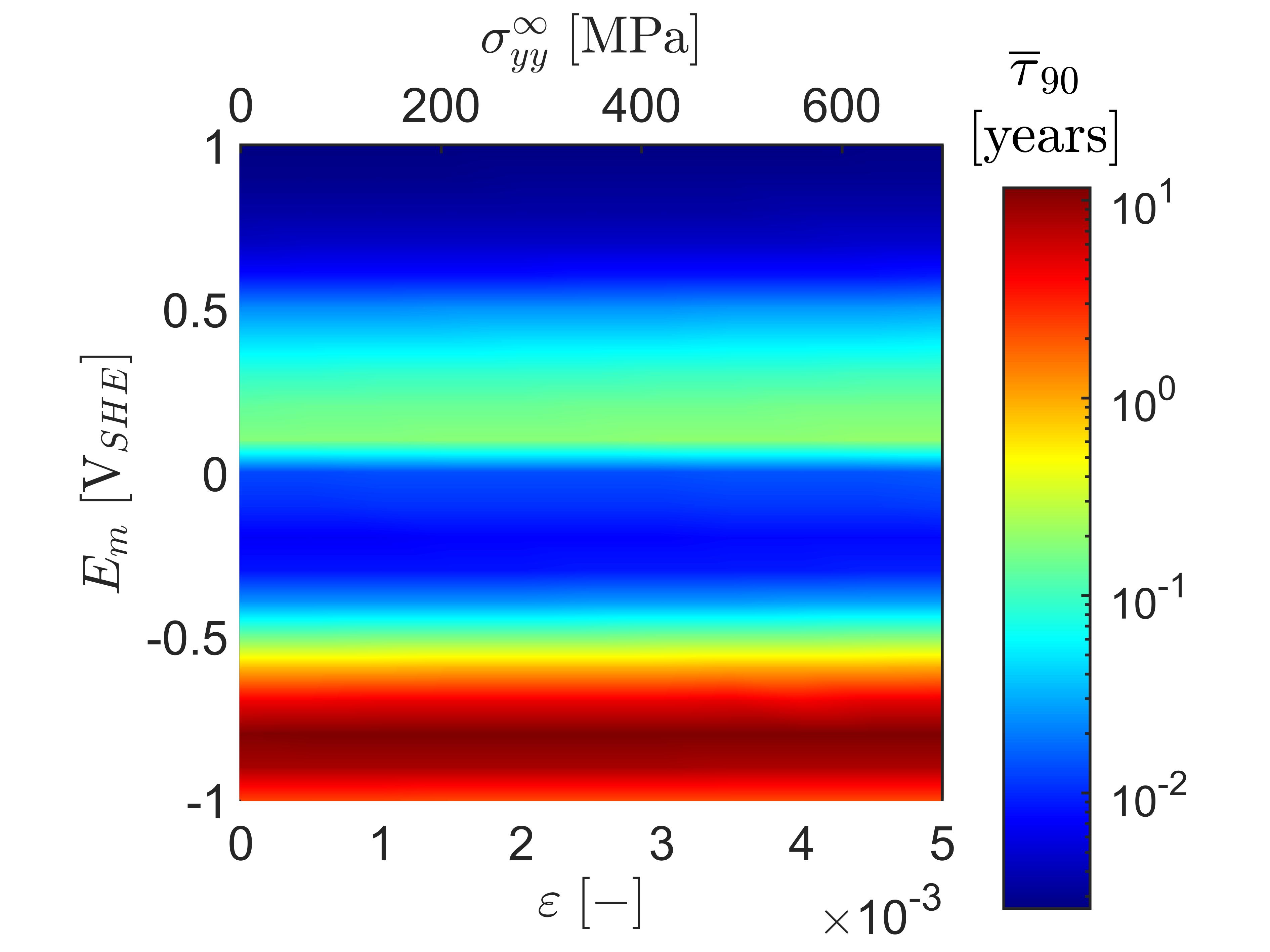}
         \caption{}
         \label{fig:rounded_t90_Average}
    \end{subfigure}
    \begin{subfigure}{7.5cm}
         \centering
         \includegraphics[width=7.5cm]{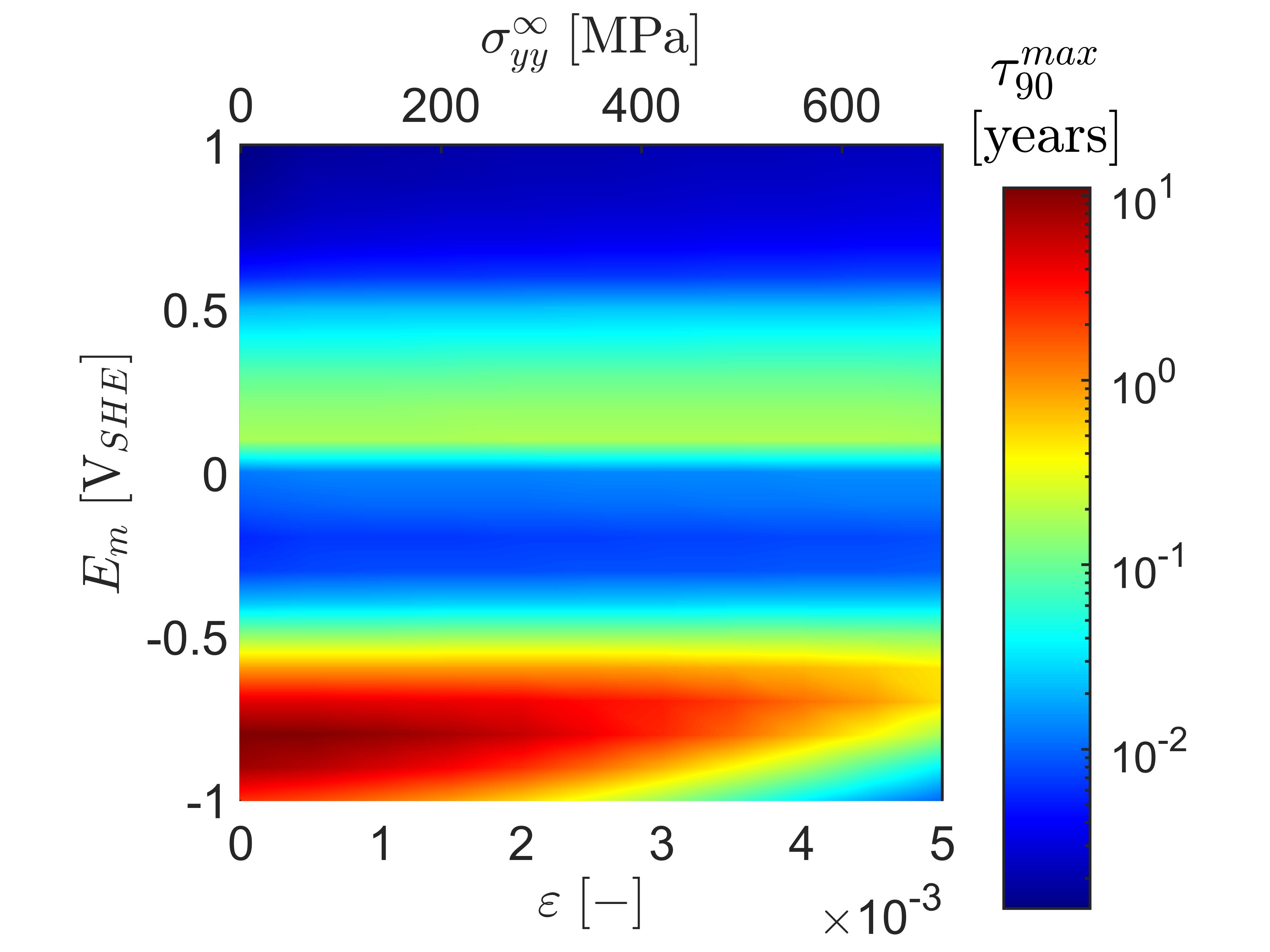}
         \caption{}
         \label{fig:rounded_t90_max}
     \end{subfigure}
    \caption{Hydrogen uptake sensitivity to mechanical straining. Time required to attain 90\% of the steady state values of (a) the average lattice hydrogen concentration $\overline{C}_L$, and (b) the maximum hydrogen concentration $C_L^{max}$. The results have been obtained for the blunted crack case, as a function of the remote strain/stress and the applied potential $E_m$.}
    \label{fig:rounded_t90}
\end{figure}
%\begin{figure}
%    \centering
%    \begin{subfigure}{8cm}
%         \centering
%         \includegraphics[width=8cm]{Figures/Rounded_surf_EPot.jpg}
%         \caption{}
%         \label{fig:rounded_environment_pH}
%    \end{subfigure}
%    \begin{subfigure}{8cm}
%         \centering
%         \includegraphics[width=8cm]{Figures/Rounded_surf_pH.jpg}
%         \caption{}
%         \label{fig:rounded_environment_EPot}
%     \end{subfigure}
%    \caption{pH (a) and electrolyte potential (b) at the crack tip for the rounded crack case under varying metal potentials and imposed strains.}
%    \label{fig:rounded_environment}
%\end{figure}
\begin{figure}
    \centering
    \begin{subfigure}{7.5cm}
         \centering
         \includegraphics[width=7.5cm]{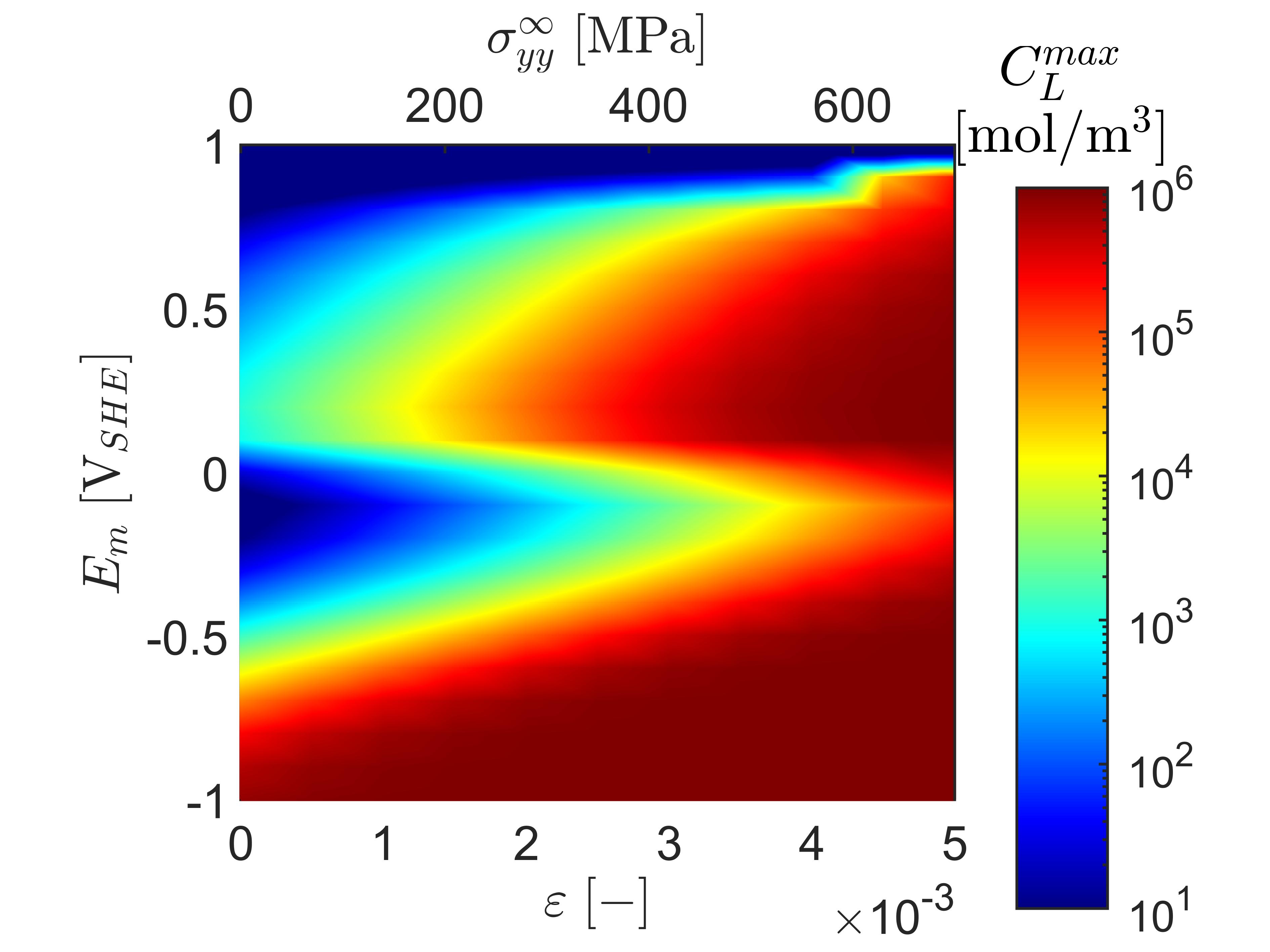}
         \caption{}
         \label{fig:Sharp_CL_Max}
    \end{subfigure}
    \begin{subfigure}{7.5cm}
         \centering
         \includegraphics[width=7.5cm]{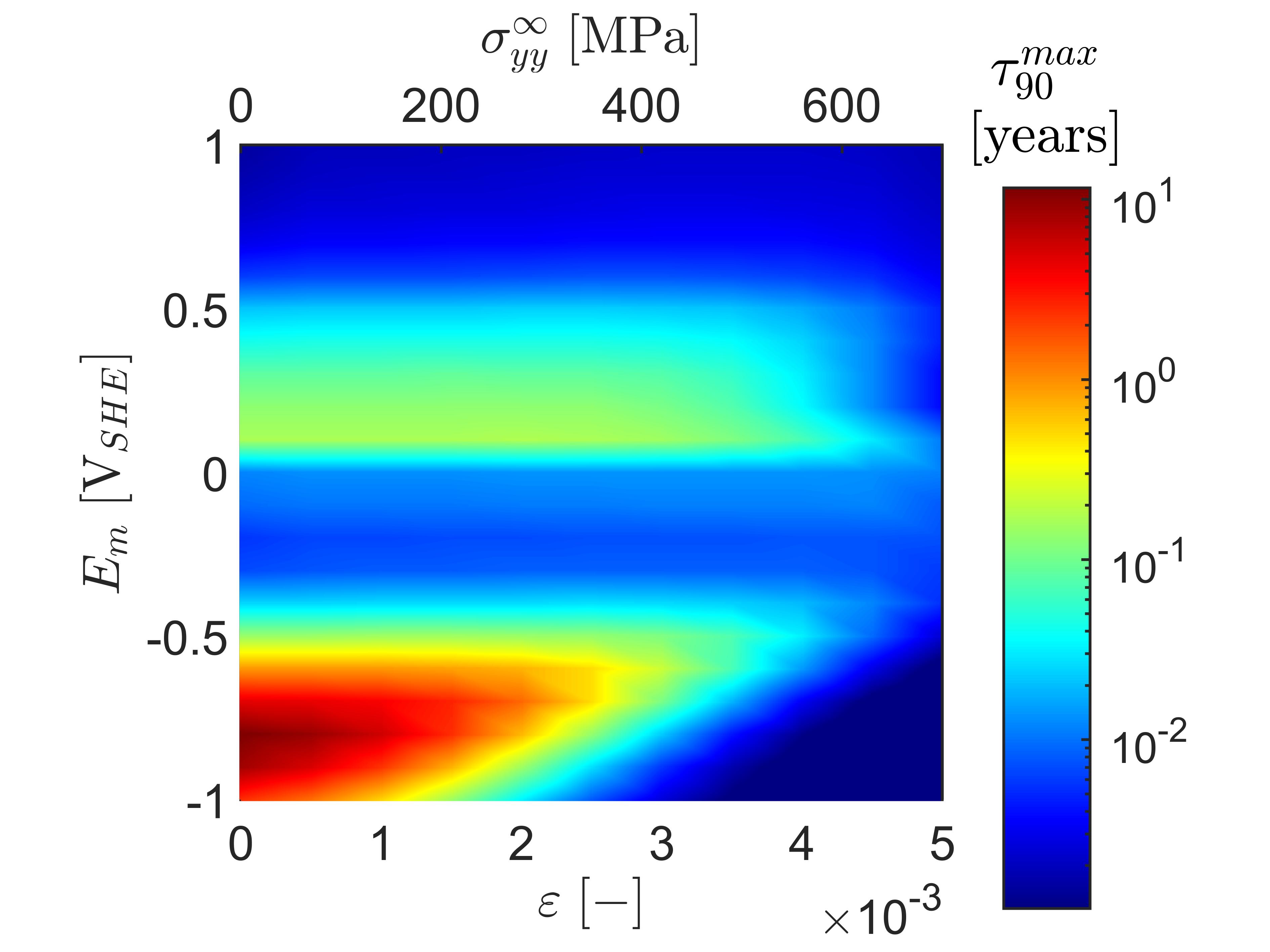}
         \caption{}
         \label{fig:Sharp_t90_max}
     \end{subfigure}
    \caption{Hydrogen uptake sensitivity to mechanical straining. Results obtained for the sharp crack case as a function of the applied load and potential, showing: (a) the maximum lattice hydrogen concentration $C_L^{max}$ at steady state, and (b) the time required to attain $90\%$ of this steady state $C_L^{max}$.}
    \label{fig:Sharp_Results}
\end{figure}

Finally, the last case study aims at shedding light into the interplay between applied mechanical stresses and hydrogen uptake, and at demonstrating the performance of the lumped integration scheme presented for relevant regimes of mechanical load. To this end, a tensile rod subjected to a prescribed remote strain is considered, see \cref{fig:domain_3D}. The metal domain has a $1\;\mathrm{cm}$ radius and $5\;\mathrm{cm}$ length, and is contained within an electrolyte domain with an outer radius of $5\;\mathrm{cm}$. The metallic sample contains a defect that starts in the outer surface and penetrates up to a depth of $5\;\mathrm{mm}$. As shown in \cref{fig:domain_3D}, the role of the defect geometry is investigated, considering both a rounded notch with radius 0.4 mm and the case of a sharp crack, with an outer opening of 0.8 mm and inner tip radius 0.05 mm. Stress concentrators result in high gradients of hydrostatic stress, which lead to an accumulation of lattice hydrogen in their vicinity. These high lattice hydrogen concentrations shift the equilibrium of the adsorption reaction, increasing the surface coverage which in turn accelerates hydrogen recombination through the Heyrovsky and Tafel reactions. The material properties of the rod and the reaction constants for the electrolyte correspond to those used previously, and are given in \cref{table:params,tab:reactionsused}. The electrolyte boundary conditions mimic those of the previous analysis; that is, we prescribe at its exterior boundary a constant concentration of the relevant ionic species and a constant electrolyte potential, using identical magnitudes to those reported in the previous study. The displacement is constrained at the bottom of the metal, while the vertical displacement at the top is prescribed based on the imposed average strain, $U_{ext}=\epsilon_{ext}\cdot 5\;\mathrm{cm}$. This strain is varied between $\epsilon_{ext}=0$ and $\epsilon_{ext}=5\cdot10^{-3}$, resulting in tensile stresses near the top and bottom of the domain of up to $\sigma_{yy} = 700\; \mathrm{MPa}$ for the used geometries. The electric potential of the metal is varied between $-1\;\mathrm{V}_{\mathrm{SHE}}$ and $1\;\mathrm{V}_{\mathrm{SHE}}$. The finite element model exploits axial symmetry, allowing the three-dimensional domain to be discretised using a two-dimensional mesh, with this mesh using a minimum element size of $0.1\;\mathrm{mm}$ near the metal-electrolyte interface and larger elements with a size of up to $1\;\mathrm{mm}$ away from this interface. Near the notch tip, the rounded notch case uses an element with a size of $10 \; \mathrm{\mu}\mathrm{m}$, while this characteristic element size equals $1 \; \mathrm{\mu}\mathrm{m}$ for the sharp crack. These meshes result in a total number of DOFs between $1.6\cdot10^5$ (blunted crack) and $2.0\cdot10^5$ (sharp crack). Details about the alterations to the previously described scheme due to the axisymmetric nature of the boundary value problem are given in \cref{sec:axisymmetric_changes}.\\ 

The steady state results for the blunted crack case are given in \cref{fig:rounded_CL}, in terms of maps for the average lattice concentration $\overline{C}_L$ (\cref{fig:rounded_CL}a) and the maximum lattice concentration $C_L^{max}$ (\cref{fig:rounded_CL}b), as a function of the applied potential $E_m$ and the remote strain/stress. The maximum hydrogen attained is highly sensitive to the applied load as crack tip stresses increase with the applied strain and this leads to higher hydrogen contents - see \cref{eq:CMAX}. However, as shown in \cref{fig:rounded_CL}a, since this effect is localised at the crack tip, the impact on the domain-wide average hydrogen concentration is negligible. The results of \cref{fig:rounded_CL}a also reveal a $\overline{C}_L$ sensitive with applied potential that qualitative resembles the findings of Fig. \ref{fig:Em_vs_CLmax}. The time required to achieve $90\%$ of the steady state concentration is shown in Figs. \ref{fig:rounded_t90}a and \cref{fig:rounded_t90}b, for the average and maximum lattice hydrogen concentrations, respectively. Since the average concentration is almost insensitive to the imposed strain, so is the time required to attain this average. In contrast, the time required to attain a $C_L^{max}$ level that is 90\% of that at steady state is dependent on the remote load. When the metal lattice is nearly saturated, increasing the strain facilitates achieving the maximum concentration faster, for instance altering the time from $3\;\mathrm{years}$ to $5\;\mathrm{days}$ when the metal potential is $-1\;\mathrm{V}_{\mathrm{SHE}}$.\\ 

The results for the sharp crack case are given in \cref{fig:Sharp_Results}, showing the maximum lattice concentration and time required to obtain this maximum concentration. Due to the increased stresses at the tip of the sharp crack, the hydrostatic stress gradient is much higher for this case compared to the rounded case. As a result, the metal lattice is saturated with hydrogen for less negative metal potentials, and for lower applied strains. Similar to the rounded case, once the metal is saturated, increasing the strains further reduces the time required to obtain the maximum concentration. Due to the larger hydrostatic stress gradient, these maxima are achieved faster, taking less than a day to obtain the equilibrium for the largest strain and lowest potential case. These results not only demonstrate the stability of the presented scheme and its ability to simulate more complex conditions, but also highlight the importance of coupled electro-chemo-mechanical effects in the prediction of hydrogen ingress. Even though the steady state average lattice hydrogen concentration does not depend on the applied loading, the concentration at the tip of the  defect depends strongly on the strain level and the defect geometry. Furthermore, the time scale over which this hydrogen enters the metal also depends on these factors, with the time required to achieve steady state conditions ranging from several days to over ten years. The results provided bring new physical insight, given the challenges associated measuring hydrogen content locally near crack tips. While this also hinders direct validation, other outputs of the model (e.g., pH) can be validated against artificial crevice electrochemical cell measurements (see Refs. \cite{Hageman2022,Gangloff2014}).

\FloatBarrier
\section*{Conclusions}
\label{sec:conclusions}

We have presented a new lumped integration scheme for reducing oscillations in the modelling of metal-electrolyte reactions. The robustness and capabilities of the scheme are demonstrated by addressing the paradigmatic case of hydrogen uptake in metals, a technologically-relevant phenomenon whose prediction is compromised by instabilities resulting from differences in reaction rates. Thus, the lumped integration scheme presented is coupled to a numerical framework resolving electrolyte behaviour, interface reactions and hydrogen bulk diffusion in the presence of a mechanical load. The simulation of several boundary value problems of particular interest reveals that:
\begin{itemize}
    \item The use of conventional Gauss integration schemes results in severe oscillations, decreasing the time increment allowed to retain stable simulations and limiting the range of parameters and time scales that can be simulated.
    \item The lumped integration scheme presented eliminates numerical oscillations and leads to smooth solutions. Furthermore, lumped integration leads to a significant improvement in tangential matrix conditioning for large time increments and high reaction rates. 
    \item The use of a lumped integration scheme enables the simulation of hydrogen uptake in metals over a wide range of material and environmental parameters. Moreover, it also enables the use of large time increments, which is essential to deliver predictions over technologically-relevant time scales. 
    \item The lumped integration scheme presented is shown to be stable over all relevant values of applied potential and reaction rate constants. Among others, this enables determining the time required to achieve steady state conditions for arbitrary choices of environment-material system.
    \item The results obtained for the tensile rod case study show that the scheme also remains stable in the presence of varying mechanical load. As a result, steady state times for hydrogen uptake near crack tips have been obtained for the first time. 
\end{itemize}

The scheme presented has shown to make feasible the simulation of hydrogen ingress in a coupled electrolyte-metal domain and could be readily adopted to tackle other relevant electrochemical systems. 

\section*{Acknowledgments}
\noindent The authors gratefully acknowledge financial support through grant EP/V009680/1 (``NEXTGEM") from the Engineering and Physical Sciences Research Council (EPSRC) and the computational resources and support provided by the Imperial College Research Computing Service (http://doi.org/10.14469/hpc/2232). Emilio Mart\'{\i}nez-Pa\~neda additionally acknowledges financial support from UKRI's Future Leaders Fellowship programme [grant MR/V024124/1].

\section*{Data availability}
\noindent The \texttt{MATLAB} code used to produce the results presented in this paper, together with documentation detailing the use of this code, are made freely available at \url{www.imperial.ac.uk/mechanics-materials/codes} and \url{www.empaneda.com}. Documentation is also provided, along with example files that enable reproduction of the results shown within the two-dimensional cases.

\FloatBarrier
\appendix

\section{Solution scheme and tangential matrices}
\label{sec:matrices}
The discretised equations, \crefrange{eq:f_u}{eq:f_theta}, are solved using a monolithic Newton-Raphson scheme. Thus, the system of governing equations,
\begin{equation}
    \begin{bmatrix} 
    \bm{K}_{uu} & \bm{0} & \bm{0} & \bm{0} & \bm{0}\\
    \bm{K}_{Lu} & \bm{K}_{LL} & \bm{K}_{L\theta} & \bm{0} & \bm{0}\\
    \bm{0} & \bm{K}_{\theta L} & \bm{K}_{\theta\theta} & \bm{K}_{\theta C} & \bm{K}_{\theta\varphi}\\
    \bm{0} & \bm{0} & \bm{K}_{C\theta} & \bm{K}_{CC} & \bm{K}_{C\varphi}\\
    \bm{0} & \bm{0} & \bm{0} & \bm{K}_{\varphi C} & \bm{0}\\
    \end{bmatrix}_i
    \begin{bmatrix} 
    \delta\mathbf{u}\\
    \delta\mathbf{C}_L\\
    \delta\bm{\uptheta}\\
    \delta\mathbf{C}_\pi \\
    \delta\bm{\upvarphi}
    \end{bmatrix}_{i+1} 
    = - 
    \begin{bmatrix} 
    \mathbf{f}_u\\
    \mathbf{f}_L\\
    \mathbf{f}_{\theta}\\
    \mathbf{f}_{C\pi} \\
    \mathbf{f}_\varphi 
    \end{bmatrix}_i
    \label{eq:Newton_Raphson}
\end{equation}
is iteratively solved, such that the iterative increment in the solution (e.g., $\delta \mathbf{u}_{i+1}$) is added to the state vector at the end of each iteration. Here, the forces $\mathbf{f}$ and tangential matrices $\bm{K}$ are updated at the start of each iteration $i$ using the then-available solution state (e.g., $\mathbf{u}_i$). The expressions for the stiffness matrices are provided below, categorised as a function of the relevant domain (metal, electrolyte, metal-electrolyte interface). 

\subsection*{A.1 Metal sub-domain}

For the momentum balance, the definition of the force vector $\mathbf{f}_u$ is given in \cref{eq:f_u}, and its associated tangential matrix is:
\begin{fleqn}\begin{equation}
    \bm{K}_{uu} = \int_{\Omega_m} \bm{B}_u^T\bm{D}\bm{B}_u \;\mathrm{d}\Omega_m
\end{equation}\end{fleqn}
Similarly, the force vector for the lattice hydrogen mass balance is given in \cref{eq:f_c}, and the relevant tangential matrices are given by
\begin{fleqn} \begin{equation}
    \bm{K}_{Lu}  = - \int_{\Omega_m} \frac{E}{3(1-2\nu)}\frac{D_L \overline{V}_H}{RT} \left(\bm{\nabla}\mathbf{N}_L\right)^T \left(\mathbf{N}_L \mathbf{C}_{L}^{t+\Delta t} \right)\bm{B}_u^*  \; \mathrm{d}\Omega_m
\end{equation}\end{fleqn} \begin{fleqn} \begin{equation}
    \begin{split}\bm{K}_{LL}  = &- \int_{\Omega_m} \frac{E}{3(1-2\nu)}\frac{D_L \overline{V}_H}{RT} \left(\bm{\nabla}\mathbf{N}_L\right)^T \left(\bm{B}_u^* \mathbf{u}^{t+\Delta t}\right) \mathbf{N}_L \; \mathrm{d}\Omega_m + \int_{\Omega_m} \frac{1}{\Delta t} \mathbf{N}_L^T\mathbf{N}_L\;\mathrm{d}\Omega_m \\
    &+  \int_{\Omega_m} \frac{D_L}{1-\mathbf{N}_L \bm{C}_L^{t+\Delta t}/N_L}\left(\bm{\nabla}\mathbf{N}_L\right)^T\bm{\nabla}\mathbf{N}_L +\frac{D_L \bm{\nabla}\mathbf{N}_L \mathbf{C}_{L\;i}^{t+\Delta t}}{N_L\left(1-\mathbf{N}_L \bm{C}_{L}^{t+\Delta t}/N_L\right)^2}\left(\bm{\nabla}\mathbf{N}_L\right)^T\mathbf{N}_L\;\mathrm{d}\Omega_{m} \\
    &- \sum_{el} \sum_{nd} \bm{W}(nd) \Big( -k_A \bm{\uptheta}_{ads}^{t+\Delta t}(nd) - k_A' (1-\bm{\uptheta}_{ads}^{t+\Delta t}(nd)\;)\Big) \bm{I}(nd,C_L) 
    \end{split}
\end{equation}\end{fleqn} \begin{fleqn} \begin{equation}
    \bm{K}_{L\theta}  = -\sum_{el} \sum_{nd} \bm{W}(nd) \Big( k_A (N_L - \mathbf{C}_L^{t+\Delta t}(nd)\;) + k_A' \mathbf{C}_L^{t+\Delta t}(nd) \Big) \bm{I}(nd,\theta)
\end{equation}\end{fleqn}
where the matrix $\bm{I}(nd,C_L)$ is used to assign the lumped integration terms to the location within the tangential sub-matrices associated with the node $nd$ and degree of freedom $C_L$.

\subsection*{A.2 Electrolyte sub-domain}

The force vectors for the mass balance and electroneutrality condition within the electrolyte are given by \cref{eq:f_c,eq:f_phi}. The related tangential matrix terms are given as follows. Firstly, for the mass balance one obtains:
\begin{fleqn}\begin{equation}
    \bm{K}_{C_{\mathrm{H}^+}\theta} =   \sum_{iel} \sum_{nd} \bm{W}(nd) \left( \frac{\partial \nu_{Va}}{\partial \theta} - \frac{\partial \nu_{Va}'}{\partial \theta} + \frac{\partial \nu_{Ha}}{\partial \theta} \right)   \bm{I}(nd,\theta)
\end{equation}\end{fleqn} \begin{fleqn} \begin{equation}
    \bm{K}_{C_{\mathrm{OH}^-}\theta} =  -\sum_{iel} \sum_{nd} \bm{W}(nd) \left( \frac{\partial \nu_{Vb}}{\partial \theta} - \frac{\partial \nu_{Vb}'}{\partial \theta} + \frac{\partial \nu_{Hb}}{\partial \theta} \right)   \bm{I}(nd,\theta) 
\end{equation}\end{fleqn} \begin{fleqn} \begin{equation}
    \begin{split}
    \bm{K}_{C_{\mathrm{H}^+}C_{\mathrm{H}^+}} = & \bm{H}_{C_{\mathrm{H}^+} C_{\mathrm{H}^+}} +\sum_{iel} \sum_{nd} \bm{W}(nd) \left( \frac{\partial \nu_{Va}}{\partial C_{\mathrm{H}^+}} + \frac{\partial \nu_{Ha}}{\partial C_{\mathrm{H}^+} } \right)   \bm{I}(nd,C_{\mathrm{H}^+}) \\ & + \sum_{el} \sum_{nd} \mathbf{W}(nd)  k_{eq} \mathbf{C}_{\mathrm{OH}^-}(nd) \bm{I}(nd,C_{\mathrm{H}^+})  + \sum_{el} \sum_{nd} \mathbf{W}(nd) k_{fe}' \mathbf{C}_{\mathrm{FeOH}^{+}}(nd) \bm{I}(nd,C_{\mathrm{H}^+})
    \end{split}
\end{equation}\end{fleqn} \begin{fleqn} \begin{equation}
    \begin{split}
    \bm{K}_{C_{\mathrm{OH}^-}C_{\mathrm{OH}^-}} = & \bm{H}_{C_{\mathrm{OH}^-} C_{\mathrm{OH}^-}} - \sum_{iel} \sum_{nd} \bm{W}(nd) \frac{\partial \nu_{Vb}'}{\partial C_{\mathrm{OH}^-}} \bm{I}(nd,C_{\mathrm{OH}^-}) \\ & + \sum_{el} \sum_{nd} \mathbf{W}(nd) k_{eq} \mathbf{C}_{\mathrm{H}^+}(nd)  \bm{I}(nd,C_{\mathrm{OH}^-})
    \end{split}
\end{equation}\end{fleqn} \begin{fleqn} \begin{equation}
    \bm{K}_{C_{\mathrm{Na}^+}C_{\mathrm{Na}^+}} = \bm{K}_{C_{\mathrm{Cl}^-}C_{\mathrm{Cl}^-}} = \bm{H}_{C_\pi C_\pi} 
\end{equation}\end{fleqn} \begin{fleqn} \begin{equation}
    \begin{split}
    \bm{K}_{C_{\mathrm{Fe}^{2+}}C_{\mathrm{Fe}^{2+}}} = & \bm{H}_{C_{\mathrm{Fe}^{2+}}C_{\mathrm{Fe}^{2+}}} \\ +& \sum_{iel} \sum_{nd} \bm{W}(nd) \left( k_c e^{-\alpha_c \left(E_m-\bm{\upvarphi}(nd)-E_{eq,c}\right) } \right)   \bm{I}(nd,C_{\mathrm{Fe}^{2+}}) + \sum_{el} \sum_{nd} \mathbf{W}(nd) k_{fe} \bm{I}(nd,C_{\mathrm{Fe}^{2+}}) \end{split}
\end{equation}\end{fleqn} \begin{fleqn} \begin{equation}
    \bm{K}_{C_{\mathrm{FeOH}^+}C_{\mathrm{FeOH}^+}} =  \bm{H}_{C_{\mathrm{FeOH}^+}C_{\mathrm{FeOH}^+}} + \sum_{el} \sum_{nd} \mathbf{W}(nd) \left(k_{fe}' \mathbf{C}_{\mathrm{H}^{+}}(nd)+k_{feoh}\right) \bm{I}(nd,C_{\mathrm{FeOH}^+}) 
\end{equation}\end{fleqn} \begin{fleqn} \begin{equation}
    \bm{K}_{C_{\mathrm{H}^+}C_{\mathrm{OH}^-}} =  \sum_{el} \sum_{nd} \mathbf{W}(nd) k_{eq} \mathbf{C}_{\mathrm{H}^+}(nd) \bm{I}(nd,C_{\mathrm{OH}^-}) 
\end{equation}\end{fleqn} \begin{fleqn} \begin{equation}
    \bm{K}_{C_{\mathrm{OH}^-}C_{\mathrm{H}^+}} =  \sum_{el} \sum_{nd} \mathbf{W}(nd) k_{eq} \mathbf{C}_{\mathrm{OH}^-}(nd) \bm{I}(nd,C_{\mathrm{H}^+})
\end{equation}\end{fleqn} \begin{fleqn} \begin{equation}
    \bm{K}_{C_{\mathrm{H}^+}C_{\mathrm{Fe}^{2+}}} =  - \sum_{el} \sum_{nd} \mathbf{W}(nd) k_{fe} \bm{I}(nd,C_{\mathrm{Fe}^{2+}})
\end{equation}\end{fleqn} \begin{fleqn} \begin{equation}
    \bm{K}_{C_{\mathrm{H}^+}C_{\mathrm{FeOH}^{+}}} =   \sum_{el} \sum_{nd} \mathbf{W}(nd) \left(k_{fe}' \mathbf{C}_{\mathrm{H}^{+}}(nd)-k_{feoh}\right) \bm{I}(nd,C_{\mathrm{FeOH}^+})
\end{equation}\end{fleqn} \begin{fleqn} \begin{equation}
    \bm{K}_{C_{\mathrm{Fe}^{2+}}C_{\mathrm{H}^{+}}} =  - \sum_{el} \sum_{nd} \mathbf{W}(nd) k_{fe}' \mathbf{C}_{\mathrm{FeOH}^{+}}(nd) \bm{I}(nd,C_{\mathrm{H}^+}) 
\end{equation}\end{fleqn} \begin{fleqn} \begin{equation}
    \bm{K}_{C_{\mathrm{Fe}^{2+}}C_{\mathrm{FeOH}^{+}}} =  - \sum_{el} \sum_{nd} \mathbf{W}(nd) k_{fe}' \mathbf{C}_{\mathrm{H}^{+}}(nd) \bm{I}(nd,C_{\mathrm{FeOH}^+})
\end{equation}\end{fleqn} \begin{fleqn} \begin{equation}
    \bm{K}_{C_{\mathrm{FeOH}^{+}}C_{\mathrm{H}^{+}}} =   \sum_{el} \sum_{nd} \mathbf{W}(nd) k_{fe}' \mathbf{C}_{\mathrm{FeOH}^{+}}(nd) \bm{I}(nd,C_{\mathrm{H}^+})
\end{equation}\end{fleqn} \begin{fleqn} \begin{equation}
    \bm{K}_{C_{\mathrm{FeOH}^{+}}C_{\mathrm{Fe}^{2+}}} =  - \sum_{el} \sum_{nd} \mathbf{W}(nd) k_{fe} \bm{I}(nd,C_{\mathrm{Fe}^{2+}})
\end{equation}\end{fleqn} \begin{fleqn} \begin{equation}
    \bm{K}_{C_{\mathrm{H}^+}\varphi}  = \bm{H}_{C_{\mathrm{H}^+} \varphi} +\sum_{iel} \sum_{nd} \bm{W}(nd) \left( \frac{\partial \nu_{Va}}{\partial \varphi} - \frac{\partial \nu_{Va}'}{\partial \varphi} + \frac{\partial \nu_{Ha}}{\partial \varphi} \right)   \bm{I}(nd,\varphi)
\end{equation}\end{fleqn} \begin{fleqn} \begin{equation}
    \bm{K}_{C_{\mathrm{OH}^-}\varphi} = \bm{H}_{C_{\mathrm{OH}^-} \varphi} -\sum_{iel} \sum_{nd} \bm{W}(nd) \left( \frac{\partial \nu_{Vb}}{\partial \varphi} - \frac{\partial \nu_{Vb}'}{\partial \varphi} + \frac{\partial \nu_{Hb}}{\partial \varphi} \right) \bm{I}(nd,\varphi) 
\end{equation}\end{fleqn} \begin{fleqn} \begin{equation}
    \bm{K}_{C_{\mathrm{Fe}^{2+}}\varphi}  = \bm{H}_{C_{\mathrm{Fe}^{2+}} \varphi} +\sum_{iel} \sum_{nd} \bm{W}(nd) \left( \frac{\partial \nu_{c}}{\partial \varphi} - \frac{\partial \nu_{c}'}{\partial \varphi} \right) \bm{I}(nd,\varphi)
\end{equation}\end{fleqn} \begin{fleqn} \begin{equation}
    \bm{K}_{C_{\mathrm{Na}^+}\varphi}  = 
    \bm{K}_{C_{\mathrm{Cl}^-}\varphi} = 
    \bm{K}_{C_{\mathrm{FeOH}^+}\varphi} = \bm{H}_{C_\pi \varphi}
\end{equation}\end{fleqn} 
where the matrix $\bm{H}_{C_\pi C_\pi}$ is defined as:
\begin{equation}
    \bm{H}_{C_\pi C_\pi} = \int_{\Omega_{e}} \frac{1}{\Delta t}\mathbf{N}_C^T \mathbf{N}_C \;\mathrm{d}\Omega_{e} + \int_{\Omega_{e}} D_\pi \left(\bm{\nabla}\mathbf{N}_c\right)^T\bm{\nabla}\mathbf{N}_C \; \mathrm{d}\Omega_{e} 
    + \int_{\Omega_{e}} \frac{D_\pi z_\pi F}{RT} \left(\bm{\nabla}\mathbf{N}_c\right)^T  \left(\bm{\nabla}\mathbf{N}_\varphi \bm{\upvarphi}^{t+\Delta t}\right) \mathbf{N}_C\;\mathrm{d}\Omega_{e}
\end{equation}
and the matrix $\bm{H}_{C_\pi \varphi}$ reads:
\begin{fleqn}\begin{equation}
    \bm{H}_{C_\pi \varphi} = \int_{\Omega_{e}} \frac{D_\pi z_\pi F}{RT} \left(\bm{\nabla}\mathbf{N}_c\right)^T \left(\mathbf{N}_c \mathbf{C}_{\pi}^{t+\Delta t}\right) \bm{\nabla}\mathbf{N}_\varphi\;\mathrm{d}\Omega_{e}
\end{equation}\end{fleqn}

Secondly, for the electroneutrality condition, the tangential matrix terms are given by:
\begin{fleqn}\begin{equation}
    \bm{K}_{\varphi C_{\pi}} =  \int_{\Omega_{e}} z_\pi  \mathbf{N}_\varphi^T\mathbf{N}_c \; \mathrm{d}\Omega_{e}
\end{equation}\end{fleqn}

\subsection*{A.3 Metal-electrolyte interface}

Finally, the relevant tangential matrices for the metal-electrolyte interface are provided. The force vector related to the mass balance of the adsorbed hydrogen is given in \cref{eq:f_theta}. Considering the case in which a lumped integration scheme is used for \textit{all} surface reactions, the tangential matrix terms are given by:
\begin{fleqn}\begin{equation}
    \bm{K}_{\theta L} =  \sum_{iel} \sum_{nd} \bm{W}(nd) \Big( -k_A \bm{\uptheta}_{ads}^{t+\Delta t}(nd) - k_A' (1-\bm{\uptheta}_{ads}^{t+\Delta t}(nd)\;)\Big) \bm{I}(nd,C_L)
\end{equation}\end{fleqn} \begin{fleqn}\begin{equation}
    \begin{split}
    \bm{K}_{\theta \theta} = & \int_{\Gamma_{int}} \frac{N_{ads}}{\Delta t} \mathbf{N}_{\theta}^T \mathbf{N}_{\theta} \; \mathrm{d}\Gamma_{int} \\ -& \sum_{iel} \sum_{nd} \bm{W}(nd) \left( \frac{\partial \nu_{Va}}{\partial \theta} - \frac{\partial \nu_{Va}'}{\partial \theta} - \frac{\partial \nu_{Ha}}{\partial \theta} - 2 \frac{\partial \nu_{T}}{\partial \theta} - \frac{\partial \nu_{A}}{\partial \theta} + \frac{\partial \nu_{A}'}{\partial \theta}+\frac{\partial \nu_{Vb}}{\partial \theta} - \frac{\partial \nu_{Vb}'}{\partial \theta} -\frac{\partial \nu_{Hb}}{\partial \theta}\right) \bm{I}(nd,\theta) 
    \end{split}
\end{equation}\end{fleqn} \begin{fleqn}\begin{equation}
    \bm{K}_{\theta C_{\mathrm{H}^+}} = - \sum_{iel} \sum_{nd} \bm{W}(nd) \left( k_{Va} e^{-\alpha_{Va} \left(E_m-\bm{\upvarphi}^{t+\Delta t}(nd)-E_{eq,Va}\right) \frac{F}{RT}} - k_{Ha} e^{-\alpha_{Ha} \left(E_m-\bm{\upvarphi}^{t+\Delta t}(nd)-E_{eq,Ha}\right) \frac{F}{RT}} \right) \bm{I}(nd,C_{\mathrm{H}^+})
\end{equation}\end{fleqn} \begin{fleqn}\begin{equation}
    \bm{K}_{\theta C_{\mathrm{OH}^-}} =   \sum_{iel} \sum_{nd} \bm{W}(nd) k_{Vb}'\bm{\uptheta}^{t+\Delta t}(nd) e^{(1-\alpha_{Vb}) \left(E_m-\bm{\upvarphi}^{t+\Delta t}(nd)-E_{eq,Vb}\right) \frac{F}{RT}} \bm{I}(nd,C_{\mathrm{OH}^-})
\end{equation}\end{fleqn} \begin{fleqn}\begin{equation}
    \bm{K}_{\theta \varphi} =  - \sum_{iel} \sum_{nd} \bm{W}(nd) \left( \frac{\partial \nu_{Va}}{\partial \varphi} - \frac{\partial \nu_{Va}'}{\partial \varphi} - \frac{\partial \nu_{Ha}}{\partial \varphi} +\frac{\partial \nu_{Vb}}{\partial \varphi} - \frac{\partial \nu_{Vb}'}{\partial \varphi} -\frac{\partial \nu_{Hb}}{\partial \varphi}\right) \bm{I}(nd,\varphi)
\end{equation}\end{fleqn}

\section{Changes relevant to an axisymmetric coordinate system}
\label{sec:axisymmetric_changes}

Axial symmetry is exploited in the tensile rod cases to simulate the three-dimensional domain depicted in Fig. \ref{fig:domain_3D}. Thus, a domain defined in the coordinate system $(r,\theta,z)$ is instead evaluated within a two-dimensional coordinate system $\mathbf{x}=(r,z)$, assuming the results are constant in the $\theta$ direction. The main notable difference resulting from this transformation is the change to the integration scheme. For instance, the absorption reaction,  \cref{eq:AbsLI}, is integrated as follows when considering axisymmetry and lumped integration:
\begin{equation}
\begin{split}
    \int_{\Gamma_{int}} \mathbf{N}_\theta^T (\nu_{Va}-\nu_{Va}')\;\mathrm{d}\Gamma_{int}(r,\theta,z) = \int_{\Gamma_{int}} 2\pi r \mathbf{N}_\theta^T (\nu_{Va}-\nu_{Va}')\;\mathrm{d}\Gamma_{int}(r,z) \\ =\sum_{iel} \sum_{nd} \bm{W}(nd) \Big( k_A (N_L - \mathbf{C}_L(nd)\;)\bm{\uptheta}_{ads}(nd) - k_A' \mathbf{C}_L(nd) (1-\bm{\uptheta}_{ads}(nd)\;)\Big)
\end{split}
\end{equation}
with the lumped integration weight vector $\mathbf{W}$ including the effects of the axisymmetry as:
\begin{equation}
    \mathbf{W} = \int_{\Gamma_{int}} 2\pi r \mathbf{N}^T\;\mathrm{d}\Gamma_{int}(r,z) = \sum_{ip} 2\pi w_{ip} r_{ip} \mathbf{N}^T(\mathbf{x}_{ip})
\end{equation}
And we proceed similarly with the volume reactions. For example, the water auto-ionisation reaction is described by
\begin{equation}
\begin{split}
    \bm{R}_{OH^-} &= -\int_{\Omega_{e}} \mathbf{N}_c^T \left(K_w-\left(\mathbf{N}_c\mathbf{C}_{\mathrm{H}^+}^{t+\Delta t}\right)\left(\mathbf{N}_c\mathbf{C}_{\mathrm{OH}^-}^{t+\Delta t}\right)\right)\;\mathrm{d}\Omega_{e}(r,\theta,z)\\&= -\int_{\Omega_{e}} 2\pi r \mathbf{N}_c^T \left(K_w-\left(\mathbf{N}_c\mathbf{C}_{\mathrm{H}^+}^{t+\Delta t}\right)\left(\mathbf{N}_c\mathbf{C}_{\mathrm{OH}^-}^{t+\Delta t}\right)\right)\;\mathrm{d}\Omega_{e}(r,z) \\
    &= -\sum_{el} \sum_{nd} \mathbf{W}(nd) \left(K_W - \mathbf{C}_{\mathrm{H}^+}(nd) \mathbf{C}_{\mathrm{OH}^-}(nd) \right)
    \end{split}
\end{equation}
using lumped weights:
\begin{equation}
    \mathbf{W} = \int_{\Omega_{el}} \mathbf{N}^T\;\mathrm{d}\Omega_{el}(r,\theta,z) =\int_{\Omega_{el}} 2\pi r \mathbf{N}^T\;\mathrm{d}\Omega_{el}(r,z) = \sum_{ip} 2\pi r_{ip} x_{ip} \mathbf{N}^T(\mathbf{x}_{ip})
\end{equation}

Another aspect to consider is the change in the strain-displacement matrix, which now reads
\begin{equation}
    \bm{B}_u = \begin{bmatrix}
    \frac{\partial N_{u1}}{\partial r} & \frac{\partial N_{u2}}{\partial r} 
    & \cdot\cdot\cdot & 
    0 & 0 
    & \cdot\cdot\cdot\\ 
    0 & 0 & \cdot\cdot\cdot &
    \frac{\partial N_{u1}}{\partial z} & \frac{\partial N_{u2}}{\partial z} & \cdot\cdot\cdot \\ N_{u1}/r & N_{u2}/r & \cdot\cdot\cdot & 0 & 0 & \cdot\cdot\cdot \\ \frac{\partial N_{u1}}{\partial z} & \frac{\partial N_{u2}}{\partial z} & \cdot\cdot\cdot & \frac{\partial N_{u1}}{\partial r}& \frac{\partial N_{u2}}{\partial r} & \cdot\cdot\cdot
    \end{bmatrix}
\end{equation}
Accordingly, the interpolation matrix used to obtain the hydrostatic stress gradient form the displacement field, $\bm{\nabla}\sigma_H = E/(3(1-2\nu))\bm{B}_u^*\mathbf{u}$, is formulated as
\begin{equation}
    \bm{B}_u^* = \begin{bmatrix} 
    \frac{\partial^2 N_{u1}}{\partial r^2}+\frac{1}{r}\frac{\partial N_{u1}}{\partial r}-\frac{N_{u1}}{r^2} & \frac{\partial^2 N_{u2}}{\partial r^2}+\frac{1}{r}\frac{\partial N_{u2}}{\partial r}-\frac{N_{u2}}{r^2} & \cdot\cdot\cdot & \frac{\partial^2 N_{u1}}{\partial r\partial z} & \frac{\partial^2 N_{u2}}{\partial r\partial z} & \cdot \cdot \cdot \\
    \frac{\partial^2 N_{u1}}{\partial r\partial z} & \frac{\partial^2 N_{u2}}{\partial r \partial z} & \cdot\cdot\cdot & \frac{\partial^2 N_{u1}}{\partial z^2} & \frac{\partial^2 N_{u2}}{\partial z^2} & \cdot \cdot \cdot
    \end{bmatrix}
\end{equation}

\def\newblock{\ }

\end{document}